\documentclass[pdftex,twocolumn,epjc3]{svjour3}          

\RequirePackage{float}
\RequirePackage{graphics}
\RequirePackage{graphicx}
\RequirePackage{epsfig}
\RequirePackage{epstopdf}
\RequirePackage{xcolor}

\smartqed  

\RequirePackage[T1]{fontenc}
\RequirePackage[utf8]{inputenc}

\RequirePackage{mathptmx}      
\RequirePackage{flushend}
\RequirePackage[numbers,sort&compress,merge]{natbib}
\RequirePackage[colorlinks,citecolor=blue,urlcolor=blue,linkcolor=blue]{hyperref}
\RequirePackage{wasysym}

\newcommand{\be}{\begin{equation}}
\newcommand{\ee}{\end{equation}}
\newcommand{\br}{\begin{eqnarray}}
\newcommand{\bea}{\begin{eqnarray}}
\newcommand{\eea}{\end{eqnarray}}
\newcommand{\er}{\end{eqnarray}}
\newcommand{\ba}{\begin{array}}
\newcommand{\ea}{\end{array}}
\newcommand{\bi}{\begin{itemize}}
\newcommand{\ei}{\end{itemize}}
\newcommand{\bn}{\begin{enumerate}}
\newcommand{\en}{\end{enumerate}}
\newcommand{\bc}{\begin{center}}
\newcommand{\ec}{\end{center}}

\newcommand{\rs}{{\sqrt s}}
\newcommand{\mm}{{\mu^+\mu^-}}

\newcommand{\gsim}{\lower1.0ex\hbox{$\;\stackrel{\textstyle>}{\sim}\;$}}
\newcommand{\lsim}{\lower1.0ex\hbox{$\;\stackrel{\textstyle<}{\sim}\;$}}

\newcommand{\bs}{\begin{small}}
\newcommand{\es}{\end{small}}

\journalname{Eur. Phys. J. C}

\begin{document}

\title{Multi Higgs production via photon fusion at future multi-TeV muon colliders}

\author{Mauro Chiesa\thanksref{e1,addr1} \and Barbara Mele\thanksref{e2,addr2} \and Fulvio Piccinini\thanksref{e3,addr1}} 

\thankstext{e1}{e-mail: mauro.chiesa@pv.infn.it}
\thankstext{e2}{e-mail: barbara.mele@roma1.infn.it}
\thankstext{e3}{e-mail: fulvio.piccinini@pv.infn.it}

\institute{INFN, Sezione di Pavia,  Via A. Bassi 6, 27100 Pavia, Italy
\label{addr1}
\and
INFN, Sezione di Roma, c/o Dipartimento di Fisica, "Sapienza" Universit\`a di Roma, P.le Aldo Moro 2, I-00185 Rome, Italy\label{addr2}
}

\date{Received: date / Accepted: date}

\maketitle

\begin{abstract}
  Multi-TeV muon colliders  promise  an unprecedented potential for   exploring the particle-physics energy frontier, and,  at the same time,   can  probe with fantastic accuracy the precise structure 
  of the Standard Model, and in particular of the Higgs boson sector. Here
  we consider  the possibility to measure  the 
  loop-suppressed single-, double-, triple-Higgs boson production  mediated by 
  the collinear-photon scattering in the channels $\mu^+\mu^-\to\mu^+\mu^-\gamma^\ast \gamma^\ast \to \mu^+\mu^- \,[H,HH,HHH]$. We study total rates and kinematical distributions in the Standard Model, and compare them with the corresponding  vector-boson-fusion processes $V^\ast V^\ast\to H,HH,HHH$ at   muon collisions 
  center-of-mass energies 1.5 TeV$\lsim\rs\lsim 100$ TeV. Possible strategies for enhancing the  
  $\gamma\gamma\to H,HH,HHH$ signal over the dominant vector-boson-fusion production are proposed. The sensitivity of total rates to possible anomalies in the Higgs-boson self-couplings is also discussed.

  \keywords{Muon collider, Higgs boson, Higgs self-couplings, VBF, loop induced processes}
  \PACS{12.15.-y,12.15.Lk,13.66.Fg,13.66.Lm,14.80.Bn}
\end{abstract}

\section{Introduction}
The planning of future colliders able to extend   
 the LHC  potential of clarifying the origin of the Standard Model (SM) present limitations 
has recently revived the interest into a multi-TeV muon collider programme~\cite{Delahaye:2019omf,Shiltsev:2019rfl,Shiltsev:2017tjx,Budker:1969cd,Ankenbrandt:1999cta,Palmer:2014nza,Wang:2015yyh,Alexahin:2018svu,
Boscolo:2018ytm,Neuffer:2018yof}. The fantastic  advantages  of using  colliding beams made up of muons, that is point-like particles,  dramatically less radiating than electrons in a circular collider, face at the moment the difficulties of realising high intensity,
low emittance muon beams which are originating from mesons 
decays via a proton driver~\cite{Parkhomchuk:1983ua,Neuffer:1983jr,Neuffer:1986dg,Kaplan:2014xda,Adey:2015iha}.
Recent progresses~\cite{MICE:2019jkl,Bonesini:2019dyo,Long:2020wfp,Young:2020wcv} as well as possible alternatives methods to
realize high intensity muon beams~\cite{Alesini:2019tlf,Antonelli:2015nla,Boscolo:2018tlu}
 might overcome in the next few years the present limitations.

Multi-TeV muon colliders could then provide a possible new  path  
to expand the energy frontier in accelerator physics~\cite{Aime:2022flm,MuonCollider:2022xlm,Black:2022cth,Accettura:2023ked}, potentially allowing 
the direct production of new heavy states, 
 even beyond the reach of  alternative future high-energy colliders  possibly following LHC~\cite{Delahaye:2019omf,Costantini:2020stv,
 AlAli:2021let,Franceschini:2021aqd}. Many possible phenomenological implications for beyond-SM searches  have recently been  considered for multi-TeV muon colliders in~\cite{Eichten:2013ckl,Chakrabarty:2014pja,Buttazzo:2018qqp,Bandyopadhyay:2020otm,Han:2021udl,Liu:2021jyc,
Han:2020uak,Capdevilla:2021fmj,Bottaro:2021srh,Capdevilla:2020qel,Buttazzo:2020eyl,Capdevilla:2021rwo,
Chen:2021rnl,Yin:2020afe,Huang:2021nkl,Huang:2021biu,Asadi:2021gah,Bottaro:2021snn,
Bottaro:2021srh,Bandyopadhyay:2021pld,Qian:2021ihf,Homiller:2022iax,Bottaro:2022one,Franceschini:2022sxc,Lv:2022pts,
Mekala:2023diu,Li:2023tbx,Li:2023ksw,Inan:2023pva,Jueid:2023zxx,Ruhdorfer:2023uea,Guo:2023jkz,Vignaroli:2023rxr,Amarkhail:2023xsc,Belyaev:2023yym,
Barducci:2023gdc,Altmannshofer:2023uci,Sun:2023ylp,Ouazghour:2023plc,Ghosh:2023xbj,Mikulenko:2023ezx,Martinez-Martinez:2023qjt,Chigusa:2023rrz,Dermisek:2023rvv,
Cassidy:2023lwd,Ake:2023xcz,Cetinkaya:2023wgg,Liu:2023jta,Bhattacharya:2023beo,Lu:2023jlr,Bandyopadhyay:2024plc,De:2024tbo,He:2024dwh}.
 At the same time, multi-TeV muon collisions can offer a 
 brand new laboratory to test with high precision the SM predictions for processes involving high-momentum transfer~\cite{DiLuzio:2018jwd,Buttazzo:2020uzc,Liu:2023yrb}.
 This holds true especially for the electroweak (EW) sector of the SM  
 whose phenomenology is moderately affected by QCD backgrounds at lepton colliders~\footnote{PDFs effects in muon beams have been considered in~\cite{Han:2020uid,Han:2021kes}.}.
In particular, Higgs boson physics can be tested in new energy regimes~\cite{Costantini:2020stv,Han:2020pif,Buttazzo:2020uzc,Han:2021lnp,Forslund:2022xjq,Forslund:2023reu}  with  
the enhancement of extremely challenging production channels like the triple Higgs
production which is sensitive to the quartic Higgs self-coupling~\cite{Chiesa:2020awd} .

\begin{figure*}
\begin{center}     
  \includegraphics[scale=0.5]{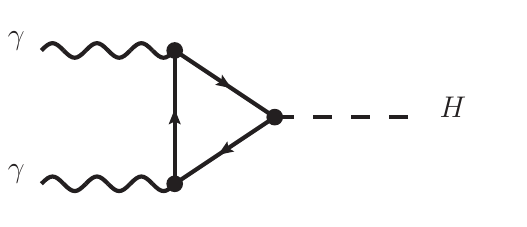}   \includegraphics[scale=0.5]{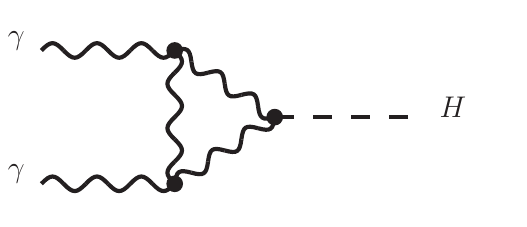}     
  \caption{\label{Feyndiags1} Feynman diagrams contributing to the process
    $\gamma\gamma\to H$, involving top-quark and $W$-boson loops.
}
   \end{center}
\end{figure*}

\begin{figure*}
\begin{center}     
 \includegraphics[scale=0.5]{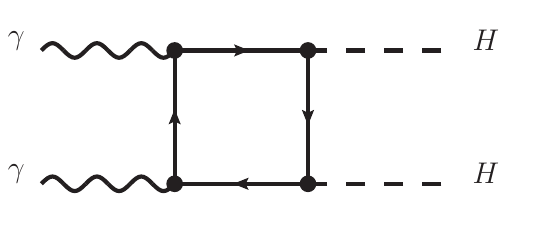}   \includegraphics[scale=0.5]{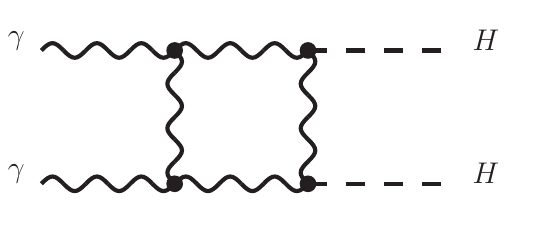} 
  \vskip0.1cm
 \includegraphics[scale=0.5]{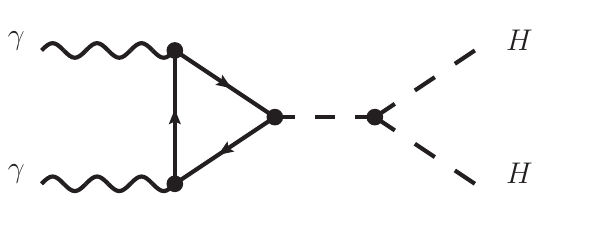}   \includegraphics[scale=0.5]{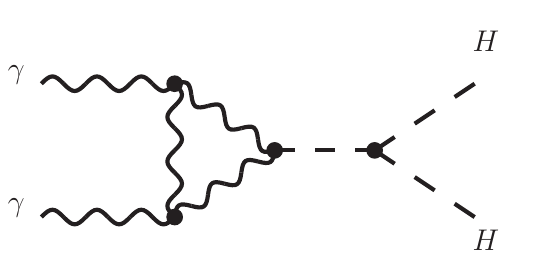} 
 \includegraphics[scale=0.5]{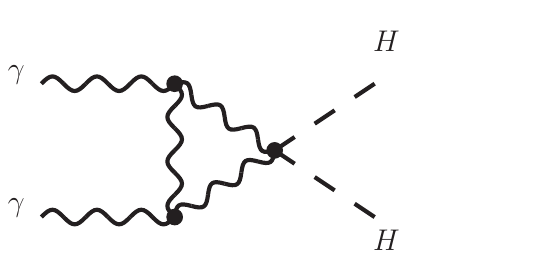} 
 \caption{\label{Feyndiags2} Feynman diagrams contributing to the process
    $\gamma\gamma\to HH$, involving top-quark and $W$-boson loops.
}
   \end{center}
\end{figure*}

\begin{figure*}
\begin{center}     
 \includegraphics[scale=0.5]{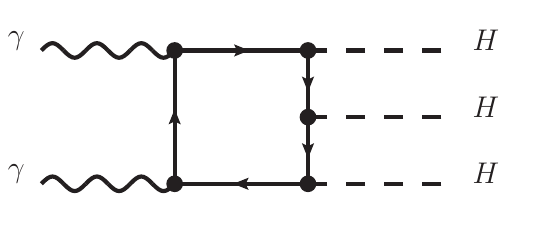}   \includegraphics[scale=0.5]{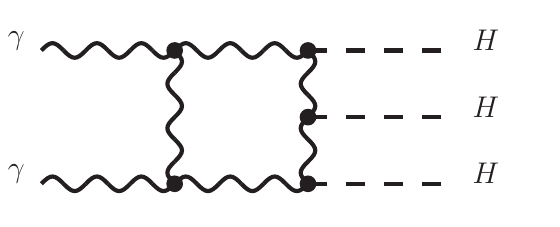} 
   \vskip0.4cm
 \includegraphics[scale=0.5]{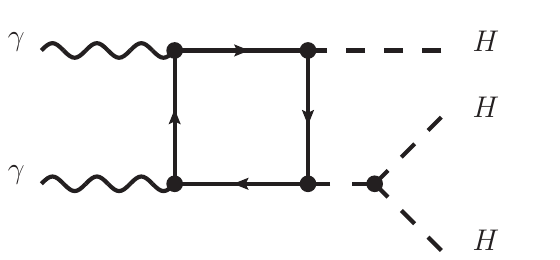}   \includegraphics[scale=0.5]{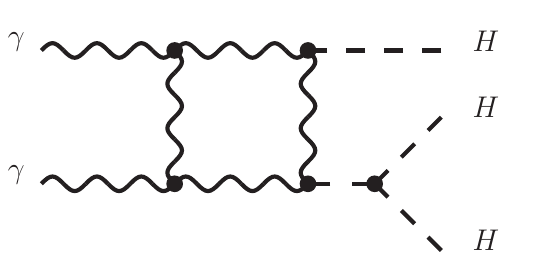}  \includegraphics[scale=0.5]{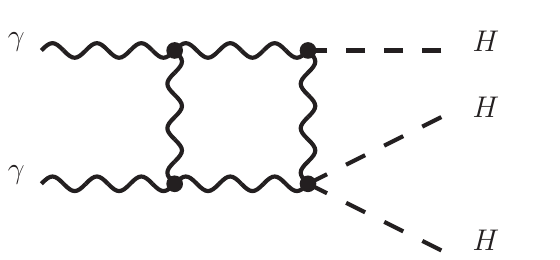} 
  \vskip0.4cm
 \includegraphics[scale=0.5]{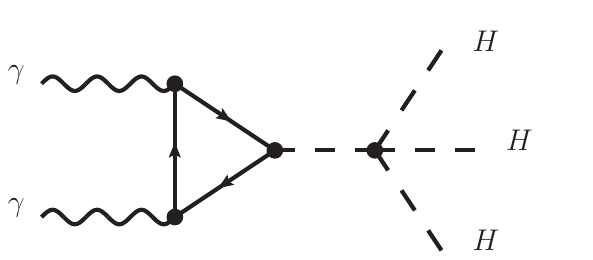}   \includegraphics[scale=0.5]{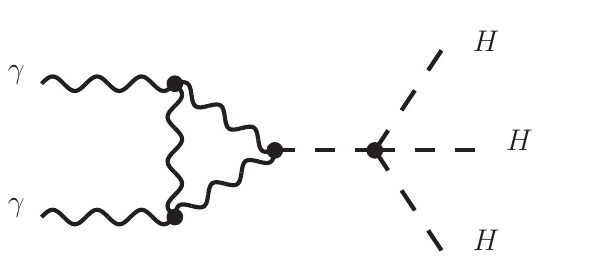}   \includegraphics[scale=0.5]{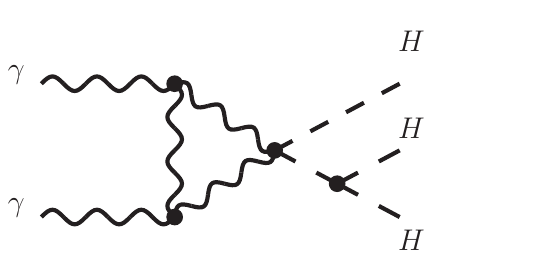} 
  \caption{\label{Feyndiags3} Representative Feynman diagrams contributing to the process
    $\gamma\gamma\to HHH$, involving top-quark and $W$-boson loops.
}
   \end{center}
\end{figure*}

Benefitting from the multi-TeV  regime, 
here we want to analyse the multiple Higgs-boson production  initiated by collinear (almost-on-shell) photons radiated 
by the initial muon beams, and mediated by  
loops of heavy particles
\begin{equation}
\mu^+\mu^-\!\to\mu^+\mu^-\gamma^\ast \gamma^\ast \to \mu^+\mu^- (n\,H)\,.
\end{equation}
In particular, we consider the production of one, two and three Higgs bosons in the channels
\bea
\mu^+\mu^- \!&\to& \mu^+\mu^- \,H\,, \\
\mu^+\mu^- \!&\to& \mu^+\mu^- \,HH\,, \\
\mu^+\mu^- \!&\to& \mu^+\mu^- \,HHH\,,
\eea
mediated by the amplitudes $\gamma^\ast \gamma^\ast \to H,HH,HHH$, respectively,  proceeding
via $W$-boson and top-quark loops.
 We show in Figures~\ref{Feyndiags1} and~\ref{Feyndiags2} the Feynman diagrams of the amplitudes $\gamma\gamma\to H,HH$, respectively,  and, in Figures~\ref{Feyndiags3}, a few representative Feynman diagrams of the amplitude $\gamma\gamma\to HHH$.
%
%
The spectrum of quasi real photons emitted by an energetic charged lepton
 beam can be described  by the leading-order  effective photon approximation (EPA)
\be
\sigma_{EPA}( \ell^{-} A \to \ell^{-} X) = \int dx\ {\cal P}_{\gamma/\ell}(x)\ \hat\sigma(\gamma A),
\ee
where, for an energy fraction $x$ of the charged lepton of initial energy $E$, the Weizs\"acker-Williams spectrum is given by \cite{vonWeizsacker:1934nji,Williams:1934ad} 
\be
{\cal P}_{\gamma,\ell}(x)  \approx   \frac{\alpha}{ 2\pi} P_{\gamma,\ell}(x) \ln{E^2\over m^2_\ell}, 
\label{eq:EPA}
\ee
with the splitting functions $P_{\gamma/\ell}(x) = (1+(1-x)^2)/x$ for $\ell\to \gamma$, 
and $P_{\ell/\ell}(x) = (1+x^2)/(1-x)$ for $\ell\to \ell$.

We  first analyze the total rates and kinematical features of the {\it partonic}
processes $\gamma \gamma \to H,HH,HHH$. We then convolute them according to the EPA 
to obtain total rates and kinematical features of the corresponding processes 
$\mu^+\mu^- \!\to \mu^+\mu^- (H,HH,HHH)\,$, mediated by collinear photons. In  EPA, the final muons will be forward and completely lost
in the detector.

Our benchmark scenarios for muon collision c.m. energies $\sqrt s$ and corresponding integrated luminosities ${\cal L}$, according to a luminosity scaling  \mbox{${\cal L}\sim  10 \;
  {\rm ab^{-1}}(\rs/10\,{\rm TeV})^2$}~\cite{Delahaye:2019omf}, will be
\be
\sqrt s \simeq 3, 6, 10, 14, 30, 100 \;{\rm TeV},
\label{energies}
\ee
and
\be
{\cal L}\sim 1, 4, 10, 20, 90, 1000 \;{\rm ab^{-1}}\, ,
\label{lumi}
\ee
respectively~\footnote{Note that more conservative (smaller) luminosity values might be more realistic in the upper
$\rs$ range  considered here (in particular for $\rs\gsim 30$ TeV).}.

In order to put the quite moderate photon-induced multi-Higgs production in context, we will compare
the $\gamma\gamma$-induced $\mu^+\mu^- \!\to \mu^+\mu^- (H,HH,HHH)\,$ rates and distributions 
 with the dominant ones for multi-Higgs production  at tree-level~\cite{Costantini:2020stv,Han:2020pif,AlAli:2021let,Buttazzo:2020uzc,Chiesa:2020awd}
\bea
\mu^+\mu^-\!&\to& \nu_\mu\bar\nu_\mu \,(H,HH,HHH)\,, \label{eq:prc1}\\
\mu^+\mu^-\!&\to& \mu^+\mu^- \!(H,HH,HHH)\,. \label{eq:prc2}
\eea
 We stress that our numerical predictions for the processes in Eqs.~(\ref{eq:prc1})--(\ref{eq:prc2})
include complete tree-level matrix elements, i.e. the contribution of both vector-boson fusion (VBF) and $s$-channel production
channels. However, since the VBF contribution is by far the dominant one at the c.m.
energies in Eq.~(\ref{energies}), we will dub the contributions in Eq.~(\ref{eq:prc1}) and Eq.~(\ref{eq:prc2})
as $WW$ and $ZZ$ fusion channels, respectively. 
The final neutrinos replacing muons in the $WW$-fusion channel 
  will then mimic the photon induced process but with different kinematical
features of the Higgs system. On the other hand, in the $ZZ$-fusion process,
the final muons  
will be  
produced at an average $p_T^{\mu}\sim M_Z$, which will differentiate the $ZZ$-fusion signal from the $\gamma\gamma$-induced one. Indeed, we will show that  in both the $WW$ and $ZZ$ fusion channels the transverse-momentum distribution for the $H,HH,HHH$ systems has a maximum at $p_T$ values of the order of the $W,Z$ masses
\be
p_T^{WW,ZZ}(H,HH,HHH)\sim M_{W,Z},
\label{ptVBF}
\ee
which is typical of vector-boson fusion induced processes (see, e.g.,~\cite{Chen:2016wkt}, and~\cite{Altarelli:1987ue} for single Higgs production in VBF).
This will give a handle to separate the latter from the 
$\gamma\gamma$-induced signal that in the EPA in Eq.~(\ref{eq:EPA})
is characterized by 
\be
p_T^{\gamma\gamma}(H,HH,HHH)\sim 0\, .
\ee
We will also compare the photon-induced loop production rates with  other $\gamma\gamma$
multi-Higgs cross sections, where the Higgs bosons are radiated at tree-level by the intermediate amplitudes
$\gamma \gamma \to W^{\scriptscriptstyle(\ast)}W^{\scriptscriptstyle(\ast)}$ and
 $\gamma \gamma \to t^{\scriptscriptstyle(\ast)}\bar{t}^{\scriptscriptstyle(\ast)}$, in the processes
\bea
\mu^+\mu^-\!&\to& \mu^+\mu^- W^+W^-(H,HH,HHH)\,, \label{eq:aaprc1}\\
\mu^+\mu^-\!&\to& \mu^+\mu^- t\bar{t} \, (H,HH,HHH)\,, \label{eq:aaprc2}
\eea
respectively,  
with forward (untagged) final muons. 
The two extra final $W$'s  and top quarks  in Eqs.~(\ref{eq:aaprc1})--(\ref{eq:aaprc2})  will of course 
distinguish these channels from the $\gamma\gamma$ loop production. We 
will see that, at particularly large $\sqrt{s}$ values, the production mechanism in Eq.~(\ref{eq:aaprc1})  becomes dominant over the  one-loop 
$\gamma\gamma$ fusion.  On the other hand, the contributions in Eq.~(\ref{eq:aaprc2}) are always smaller than the loop-induced ones
in the setups under consideration.

The plan of the paper is the following.
In Section~2, we describe the computational method used
to obtain the photon-photon one-loop amplitudes and cross sections
for multiple Higgs production.
In Section~3, we study the partonic $\gamma\gamma\to (nH)$ cross sections and kinematical distributions versus $\sqrt s_{\gamma\gamma}$.
The convoluted  $\mu^+\mu^-\!\to\mu^+\mu^-\gamma^\ast \gamma^\ast \to \mu^+\mu^- (nH)$   cross sections (kinematical distributions) are considered in Section~4~(5), and compared with the corresponding VBF processes. In Section~6, we discuss a possible strategy to enhance
the photon induced signal over the VBF production. We then study the sensitivity of the $\mu^+\mu^-\!\to\mu^+\mu^-\gamma^\ast \gamma^\ast \to \mu^+\mu^- HH$ cross section to a trilinear anomalous Higgs self-coupling, and of the $\mu^+\mu^-\!\to\mu^+\mu^-\gamma^\ast \gamma^\ast \to \mu^+\mu^- HHH$ cross section to a trilinear or quartic anomalous Higgs self-coupling in Section~7. Our conclusions are presented in Section~8.


\section{Cross section computational method}
In this section we describe the procedure we followed to obtain the amplitudes and corresponding cross sections for the different processes described below.

The matrix elements for the loop-induced processes $\gamma\gamma\to H$, $\gamma\gamma\to HH$,
and $\gamma\gamma\to HHH$ (cf. Figures~\ref{Feyndiags1}-\ref{Feyndiags3}) have been computed in a fully-automated way using the {\tt Recola}
library~\cite{Actis:2012qn,Actis:2016mpe,Denner:2017vms,Denner:2017wsf},  which relies on the {\tt Collier}
library~\cite{Denner:2010tr,Denner:2002ii,Denner:2005nn,Denner:2016kdg} for the evaluation of the tensor
and scalar one-loop integrals.

For the phase-space integration, we used the Monte Carlo integrator already employed in~\cite{Chiesa:2020ttl,Chiesa:2018lcs}
for diboson production in hadronic collisions. In the Monte Carlo program, the proton PDFs have been
replaced by a photon PDF describing the collinear $\mu \to \mu \gamma$ splitting as in Eq.~(\ref{eq:EPA}),
where $m_\ell$ is the muon mass and $E$ is the nominal energy of the muon beam.

The same procedure has been followed to compute the cross sections for 
the tree-level processes  $\gamma\gamma\to (nH)W^+W^-$ and $\gamma\gamma\to (nH) t \bar{t}$ ($n=1$, $2$, $3$),
relevant for the channels in Eqs.~(\ref{eq:aaprc1})--(\ref{eq:aaprc2}),
and for the subsequent convolution with the photon PDF in the muon beams.

The tree-level results for $\mu^+\mu^- \to (H,HH,HHH)\nu_\mu \bar{\nu}_\mu$
via $WW$ fusion, and 
$\mu^+\mu^- \to (H,HH,HHH)\mu^+\mu^-$ via $ZZ$ fusion
have been obtained by using the {\tt Whizard} Monte Carlo event 
generator~\cite{Kilian:2007gr,Moretti:2001zz}.
 In order to prevent  singularities arising from $s$-channel diagrams, we include the gauge-boson widths in a gauge-invariant way
by means of the complex-mass scheme~\cite{Denner:1999gp,Denner:2005fg,Denner:2006ic}. We checked that, for these two classes of processes, the effect of $s$-channel diagrams
is essentially dominated by the $Z$ resonance region and it is numerically relevant only for the 3~TeV setup in Eq.~(\ref{energies}),
in particular for $HHH$ production, while it becomes completely negligible when increasing the collider energy.

\section{Partonic $\gamma\gamma$ cross sections}
In this section, we discuss the total rates and kinematical distributions
for the partonic channels  
$\gamma\gamma\to H,HH,HHH$, 
 in the $\gamma\gamma$ c.m. system, before the convolution with the photon energy spectrum in the muon beams. 

In the present approximation, the single Higgs channel $\gamma\gamma\to H$
is characterized by a production rate resonating at the Higgs mass, $\sigma\sim \delta(\rs-m_H)$, with the Higgs
boson  at rest in the $\gamma\gamma$ c.m. system. Once convoluted with the 
collinear photon spectrum in muon collisions, the final Higgs will still have zero transverse momentum, and a rapidity distribution that will just reflect the one
of the initial $\gamma\gamma$ system.

\begin{figure}
\begin{center}
  \includegraphics[width=0.94\columnwidth]{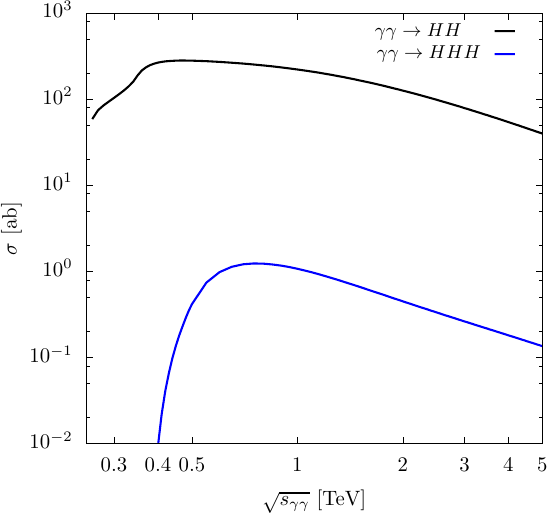}
  \caption{Partonic total cross sections for the loop-induced processes $\gamma \gamma \to HH$
  (black line), and 
  $\gamma \gamma \to HHH$ (blue line) versus the $\gamma \gamma$ c.m. energy.
\label{part1}
  }
  \end{center}
\end{figure}

\begin{figure*}
\vskip 0 cm
  \includegraphics[width=0.64\columnwidth]{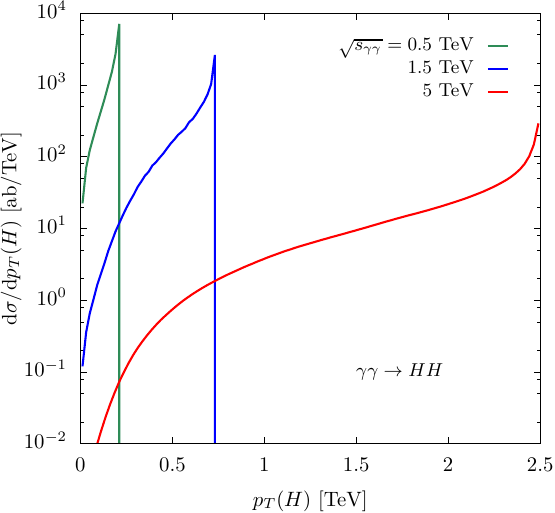}
  \hskip 0.2 cm \includegraphics[width=0.64\columnwidth]{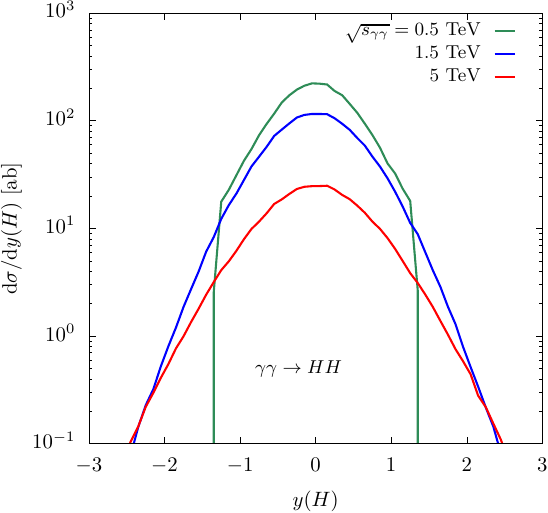} 
  \hskip 0.2 cm\includegraphics[width=0.64\columnwidth]{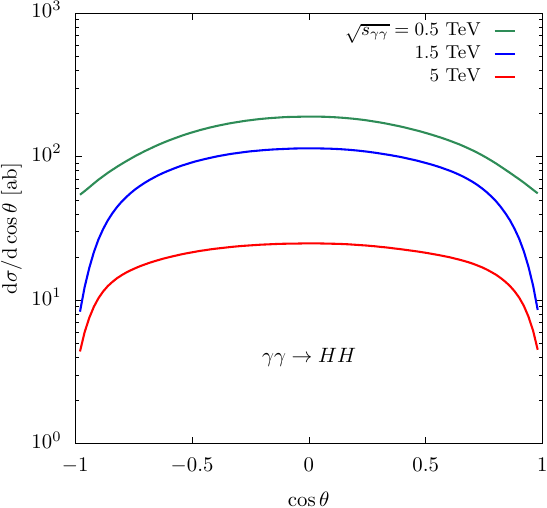}
  \caption{Higgs transverse-momentum (left), rapidity (center), and angular (right) distributions
   for the loop-induced process $\gamma \gamma \to HH$ in photon collisions,
   in the $\gamma\gamma$ 
c.m. system,
  for three values of the $\gamma \gamma$ c.m. energy.
  }
  \label{part2}
\end{figure*}

As for the partonic $\gamma\gamma\to HH,HHH$ productions, we present in Figure~\ref{part1} the corresponding total cross sections versus the $\gamma\gamma\;$ c.m. energy $\sqrt s_{\gamma\gamma}$, in the range
$2m_H\lsim$ $\sqrt s_{\gamma\gamma}\lsim 5$ TeV.
The two-Higgs cross section reaches its maximum value  
\be
\sigma_{max}(\gamma\gamma\to HH)\sim 0.28 \;{\rm fb \;\;\;\;\; for \;\;\;\;\;}
\sqrt s_{\gamma\gamma}\sim 470 \;{\rm GeV},
\ee 
while the three-Higgs cross section has a maximum
\be
\sigma_{max}(\gamma\gamma\to HHH)\sim 1.2 \;{\rm ab \;\;\;\;\; for \;\;\;\;\;}
\sqrt s_{\gamma\gamma}\sim 750 \;{\rm GeV}.
\ee 
The Higgs transverse-momentum, rapidity and angular distributions for the 
$\gamma\gamma\to HH$ process in the $\gamma\gamma$ 
c.m. system are shown in 
Figure~\ref{part2}, with the two Higgs bosons having the same  energy
$E_{H}=\sqrt s_{\gamma\gamma}/2$, and a back-to-back configuration.

\begin{figure*}
  \includegraphics[width=0.94\columnwidth]{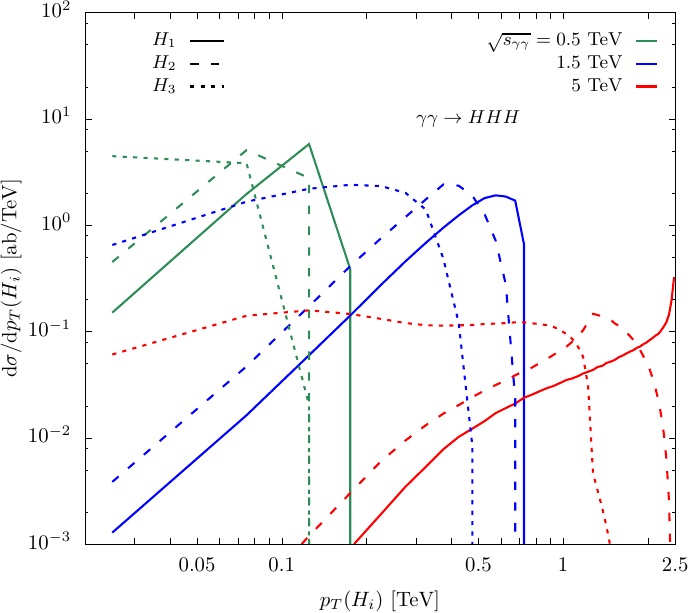} \hskip 0.6  cm\includegraphics[width=0.94\columnwidth]{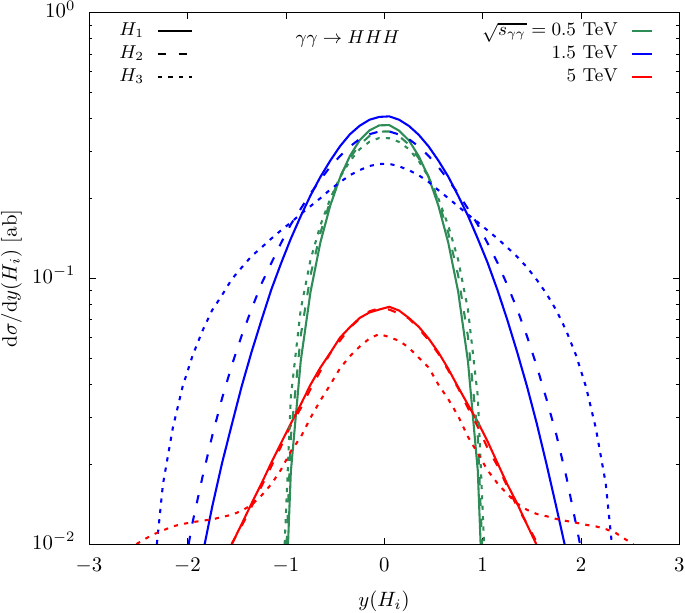}
  \vskip 0.5 cm
    \includegraphics[width=0.94\columnwidth]{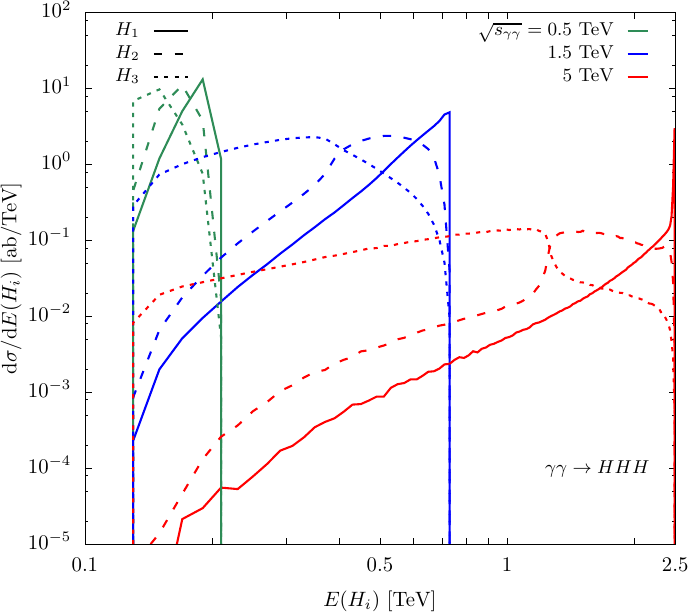}\hskip 0.7 cm\includegraphics[width=0.94\columnwidth]{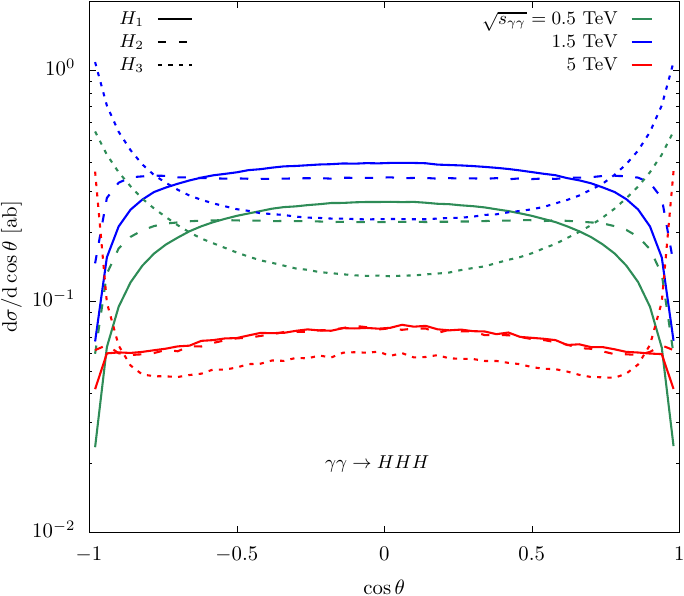}
  \caption{Higgs transverse-momentum (upper left),  rapidity  (upper right),
 energy (lower left), and angular  (lower right)  distributions
   for the loop-induced process $\gamma \gamma \to HHH$ in photon collisions,
   in the $\gamma\gamma$ 
c.m. system,
  for three values of the $\gamma \gamma$ c.m. energy.
  The  $H_i$ label ($i=1,2,3$) refers to the  Higgs bosons ordered in  $p_T$,
  with $H_1$ labelling the highest-$p_T$ Higgs.
  }
  \label{part3}
\end{figure*}

Figure~\ref{part3} shows the $\gamma\gamma\to HHH$ case (including the nontrivial Higgs energy distribution), by ordering the three final Higgs-bosons labels 
according to their transverse momentum.

\section{$\gamma\gamma$-induced $\;\mu^+\mu^- \!\to \mu^+\mu^- (H,HH,HHH)\,$
cross sections}

In this section, we present the $\mu^+\mu^-\!$ total cross sections for the $\gamma\gamma$-induced one-loop $(H,HH,HHH)$ production after the convolution
of the partonic cross section discussed in Section~3 with  the photon spectrum described by the EPA in Eq.~(\ref{eq:EPA}).

In Table~\ref{tab:xsect},
the inclusive $\mu^+\mu^-$ cross sections (in ab) are presented
 versus  $\rs$, in the range 1.5 TeV$ \lsim \rs \lsim $ 100 TeV. For comparison, $\mu^+\mu^-\!$ cross sections for other  {\it tree-level} $(H,HH,HHH)$ production  mechanisms mediated by tree-level 
   $\gamma\gamma$ fusion ($\gamma\gamma\to (nH)W^+W^-$ and  $\gamma\gamma\to (nH)t\bar{t}$, with $n=1,2,3$),
  $WW$ fusion
($WW\to H$, $WW\to HH$,
and $WW\to HHH$), and $ZZ$ fusion
($ZZ\to H$, $ZZ\to HH$,
and $ZZ\to HHH$)
 are also detailed.
 
 Regarding the single Higgs boson production, one can see that the 
  $\gamma\gamma\to H$ cross section varies in the range $\sigma_{(\gamma\gamma\to H)}\sim [0.2-1.6]$ fb. At $\rs \sim 14$ TeV, $\sigma_{(\gamma\gamma\to H)}\sim 0.78$ fb,
 which, with the benchmark integrated luminosity of 20 ab$^{-1}$ in Eq.~(\ref{lumi}),
 would correspond to about 16.000 events. At $\rs \sim 3$ TeV with 1 ab$^{-1}$,
 one  has $\sigma_{(\gamma\gamma\to H)}\sim 0.37$~fb, corresponding to  about 400 events. These rates are about 3 and 2 orders of magnitude less than,
 respectively, the
 $WW\to H$ and $ZZ\to H$ ones for all $\rs$ values.
 
  As for  double Higgs boson production,  the 
  $\gamma\gamma\to HH$ cross section varies in the range $\sigma_{(\gamma\gamma\to HH)}\sim [0.4-8]$ ab. At $\rs \sim 14$ TeV, $\sigma_{(\gamma\gamma\to HH)}\sim 3$ ab,
 which, with  20 ab$^{-1}$,
 would correspond to about 60 events. At $\rs \sim 3$ TeV with 1 ab$^{-1}$,
 one  has $\sigma_{(\gamma\gamma\to HH)}\sim 0.9$~ab, corresponding to  about 1 event.  Similarly to single Higgs production, these $HH$ rates are,
 about 3 and 2 orders of magnitude less than,
 respectively, the
 $WW\to HH$ and $ZZ\to HH$ ones for all $\rs$ values.
 
 Finally, 
for  triple Higgs boson production,  the 
  $\gamma\gamma\to HHH$ cross section varies in the range $\sigma_{(\gamma\gamma\to HHH)}\sim [0.07-2.3]\times 10^{-2}$~ab, and requires
  the maximum c.m. energy to become observables. 
    Indeed, at $\rs \sim 14$ TeV, $\sigma_{(\gamma\gamma\to HHH)}\sim 8.4\times 10^{-3}$~ab,
 which, with  20 ab$^{-1}$,
 would correspond to no event produced, just as in the case of $\rs \sim 3$ TeV (with 1 ab$^{-1}$), where 
 one  has $\sigma_{(\gamma\gamma\to HHH)}\sim 2\times 10^{-3}$ ab.
 On the other hand, at $\rs \sim 30$ TeV, $\sigma_{(\gamma\gamma\to HHH)}\sim 1.3\times 10^{-2}$~ab,
 which, with  90 ab$^{-1}$,
 would correspond to about one event.
 Hence, only going to the very upper $\rs$ range considered here, 
 one could have an observable  event number.
 I particular, at $\rs \sim 100$ TeV with 10$^3$ ab$^{-1}$, $\sigma_{(\gamma\gamma\to HHH)}\sim 2.3\times 10^{-2}$~ab,
 one would give  about 23 events.
Note that, at low $\rs$ values, the $HHH$ rates are less depleted than 
in the case
of $H$ and $HH$ production with respect
to VBF production, being only  
 about 2 and 1 orders of magnitude less 
 than the
 $WW\to HHH$ and $ZZ\to HHH$ ones,  respectively. 
 
\begin{figure*}
  \includegraphics[width=0.94\columnwidth]{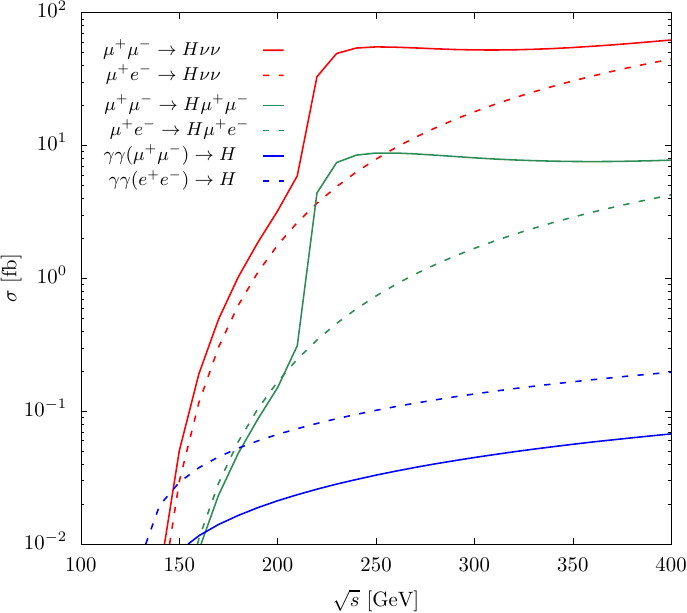}\hskip 0.3 cm\includegraphics[width=0.98\columnwidth]{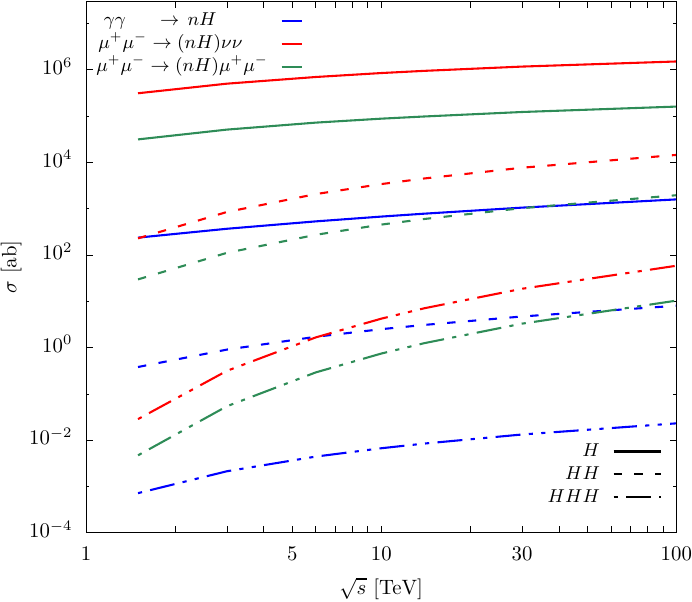}
   \caption{\label{tab:xsect2} Inclusive $\mu^+\mu^-$ cross sections  for 
   one-loop $\gamma\gamma\to H,HH,HHH$ production
   versus $\mu^+\mu^-$ c.m. energy. In the left plot the  photon-induced  single Higgs 
   $\mu^+\mu^-\to H \mu^+\mu^-$ production (blue solid line)
   is compared  with the same process  for $\mu^\pm \to e^\pm$
   (blue dashed line),    and with the tree-level processes $\mu^+\mu^-\to H \nu \bar{\nu}$ (red solid line) and $\mu^+\mu^-\to H \mu^+\mu^-$ (green solid line)  
   , for $\rs \lsim 400$ GeV.  The dashed red and green lines represent the integrated cross-sections for single Higgs productions in $WW$ and $ZZ$ fusion, respectively,
   in $\mu^+e^-$ scattering.  
   In the right plot, $\mu^+\mu^-$ cross sections for the $H$ (solid lines),
   $HH$ (dashed lines), $HHH$ (dot-dashed lines) 
   $\gamma\gamma$-mediated one-loop    production (blue lines) are compared with the corresponding total rates for 
 $WW$-fusion (red lines), and $ZZ$ fusion (green lines) production, for
 1 TeV$\lsim \rs \lsim 100$ TeV.   }
\end{figure*}

 The  behavior of $\mu^+\mu^-\!$ total rates versus $\rs$ is more 
 clearly shown in 
 Figure~\ref{tab:xsect2}, where for comparison also VBF cross sections are 
 plotted.
 In the left plot the  
   $\mu^+\mu^-\to H \mu^+\mu^-$ production  (blue solid line)
   is compared  with the same process  when replacing $\mu^\pm \to e^\pm$
   (blue dashed line), and with $WW$-fusion (red  solid  line), $ZZ$-fusion (green  solid  line)  single-Higgs production,   for $\rs \lsim 400$ GeV. The difference of about a factor three in the rates for the electron and muon  initiated processes  for $\gamma\gamma$ fusion  stems from the double logarithm arising from  Eq.~\ref{eq:EPA} applied twice to   initial beams.
    The dashed red and green lines represent the integrated cross-sections for single Higgs productions in \emph{pure} $WW$ and $ZZ$ fusion, respectively,
   in $\mu^+e^-$ scattering: in practice, since both electrons and muons in our simulations are treated as massless particles
   (with the only exception of the mass parameter in the photon spectrum in Eq.~(\ref{eq:EPA})),
   these lines correspond to the predictions for $H$ production in $WW$ and $ZZ$ fusion when only the $t$-channel production mechanism
   is considered. The difference between red (green) solid and dashed lines is large at the opening of the $Z$ resonance ($\sqrt{s}\sim M_H+M_Z$)
   and decreases for larger values of $\sqrt{s}$. \\
     In the right plot of Figure~\ref{tab:xsect2}, $\mu^+\mu^-$ collision rates for the $H$ (solid lines),
   $HH$ (dashed lines), $HHH$ (dot-dashed lines) 
    one-loop  $\gamma\gamma$ production (blue lines) are shown, and compared with the corresponding  rates for 
 tree-level $WW$-fusion (red lines), and $ZZ$ fusion (green lines) production, for
 1 TeV$\lsim \rs \lsim 100$ TeV.   In this partonic center of mass range,
 the $\mu^+\mu^-\to nH\nu \bar{\nu}$ over $\mu^+\mu^-\to nH\mu^+\mu^-$ cross-section ratio is essentially flat,
 ranging from about~10 for single-Higgs production to approximately 6 for the triple-Higgs case: this is
 mainly a consequence of the differences in the $HVV$, $HHVV$, and $Vf\overline{f}$ couplings for $V=W,Z$
 with the leading contribution coming from the latter interaction as documented in Ref.~\cite{Altarelli:1987ue}
 for single-Higgs production.

\begin{table*}
 \resizebox{\textwidth}{!}{%
  \begin{tabular}{ | l | c | c | c | c | c | c | c |}
    \hline
    $\sqrt{s}$  [TeV]       & 1.5  &  3  &  6 &  10  & 14 &  30  & 100 \\
    \hline
    $\sigma(\mu^+\mu^-\!)$     [ab]       &    &     &    &      &    &   &  \\
    \hline
    $\gamma\gamma\to H$             & 2.35$\times 10^{2}$  & 3.66$\times 10^{2}$  & 5.29$\times 10^{2}$  & 6.71$\times 10^{2}$  & 7.76$\times 10^{2}$  & 1.05$\times 10^{3}$  & 1.58$\times 10^{3}$        \\
    $\gamma\gamma \to H WW$         & 5.40$\times 10^{1}$  & 2.88$\times 10^{2}$  & 9.93$\times 10^{2}$  & 2.08$\times 10^{3}$  & 3.20$\times 10^{3}$  & 7.46$\times 10^{3}$  & 2.20$\times 10^{4}$         \\
    $\gamma\gamma \to H t\bar{t}$   & 9.53$\times 10^{-1}$ & 3.27                 & 7.53                 & 1.21$\times 10^{1}$  & 1.57$\times 10^{1}$  & 2.61$\times 10^{1}$  & 4.82$\times 10^{1}$ \\
    $\mu^+\mu^-\to H\nu \bar{\nu}$         & 3.11$\times 10^{5}$  & 4.98$\times 10^{5}$  & 6.96$\times 10^{5}$  & 8.44$\times 10^{5}$  & 9.43$\times 10^{5}$  & 1.17$\times 10^{6}$  & 1.52$\times 10^{6}$        \\    
    $\mu^+\mu^-\to H\mu^+\mu^-$     & 3.13$\times 10^{4}$  & 5.08$\times 10^{4}$  & 7.16$\times 10^{4}$  & 8.73$\times 10^{4}$  & 9.78$\times 10^{4}$  & 1.21$\times 10^{5}$  & 1.59$\times 10^{5}$       \\

    \hline
    $\gamma\gamma\to HH$            & 3.77$\times 10^{-1}$ & 9.00$\times 10^{-1}$ & 1.70                 & 2.47                 & 3.07                 & 4.67                 & 7.96           \\
    $\gamma\gamma \to HH WW$        & 5.63$\times 10^{-3}$ & 8.16$\times 10^{-2}$ & 5.72$\times 10^{-1}$ & 1.79	                & 3.43		       & 1.21$\times 10^{1}$  & 5.86$\times 10^{1}$          \\
    $\gamma\gamma \to HH t\bar{t}$  & 1.72$\times 10^{-3}$ & 9.15$\times 10^{-3}$ & 2.57$\times 10^{-2}$ & 4.55$\times 10^{-2}$ & 6.25$\times 10^{-2}$ & 1.14$\times 10^{-1}$ & 2.33$\times 10^{-1}$ \\
    $\mu^+\mu^-\to HH\nu \bar{\nu}$        & 2.43$\times 10^{2}$  & 8.48$\times 10^{2}$  & 2.05$\times 10^{3}$  & 3.38$\times 10^{3}$  & 4.47$\times 10^{3}$  & 7.60$\times 10^{3}$  & 1.44$\times 10^{4}$        \\
    $\mu^+\mu^-\to HH\mu^+\mu^-$    & 3.22$\times 10^{1}$  & 1.12$\times 10^{2}$  & 2.72$\times 10^{2}$  & 4.50$\times 10^{2}$  & 5.96$\times 10^{2}$  & 1.02$\times 10^{3}$  & 1.94$\times 10^{3}$       \\
    
    \hline
    $\gamma\gamma\to HHH  $         & 7.13$\times 10^{-4}$ & 2.11$\times 10^{-3}$ & 4.38$\times 10^{-3}$ & 6.63$\times 10^{-3}$ & 8.38$\times 10^{-3}$ & 1.31$\times 10^{-2}$ & 2.30$\times 10^{-2}$         \\
    $\gamma\gamma \to HHH WW$       & 2.57$\times 10^{-6}$ & 3.69$\times 10^{-5}$ & 2.69$\times 10^{-4}$ & 9.76$\times 10^{-4}$ & 2.15$\times 10^{-3}$ & 1.07$\times 10^{-2}$ & 8.5(1)$\times 10^{-2}$            \\
    $\gamma\gamma \to HHH t\bar{t}$ & 3.79$\times 10^{-6}$ & 4.03$\times 10^{-5}$ & 1.39$\times 10^{-4}$ & 2.67$\times 10^{-4}$ & 3.83$\times 10^{-4}$ & 7.50$\times 10^{-4}$ & 1.66$\times 10^{-3}$ \\
    $\mu^+\mu^-\to HHH\nu \bar{\nu}$       & 9.90$\times 10^{-2}$ & 3.44$\times 10^{-1}$ & 1.66	         & 4.19		        & 7.02                 & 1.85$\times 10^{1}$  & 5.82$\times 10^{1}$             \\
    $\mu^+\mu^-\to HHH\mu^+\mu^-$   & 1.66$\times 10^{-2}$ & 5.91$\times 10^{-2}$ & 2.91$\times 10^{-1}$ & 7.37$\times 10^{-1}$ & 1.24                 & 3.26                 & 1.03$\times 10^{1}$               \\
    \hline
  \end{tabular}%
  }
  \caption{\label{tab:xsect} Inclusive $\mu^+\mu^-$ cross sections (in ab) for 
  photon-induced one-loop $\gamma\gamma\to H,HH,HHH$ production 
  versus $\mu^+\mu^-$ c.m. energy. For comparison, $\mu^+\mu^-$ cross sections for other $(H,HH,HHH)$  tree-level production  mechanisms mediated by 
  $\gamma\gamma$  [$\gamma\gamma\to (nH)W^+W^-$ and $\gamma\gamma\to (nH)t\bar{t}$] and the ones for the processes
  $\mu^+\mu^-\to (nH)\nu \bar{\nu}$ and $\mu^+\mu^-\to (nH)\mu^+\mu^-$ ($n=1$, $2$, $3$)  are also detailed.}
\end{table*}

\section{Higgs boson distributions in  
$\gamma\gamma$-induced $HH\,$ and $HHH\,$ production
}
\begin{figure*}
  \includegraphics[width=0.90\columnwidth]{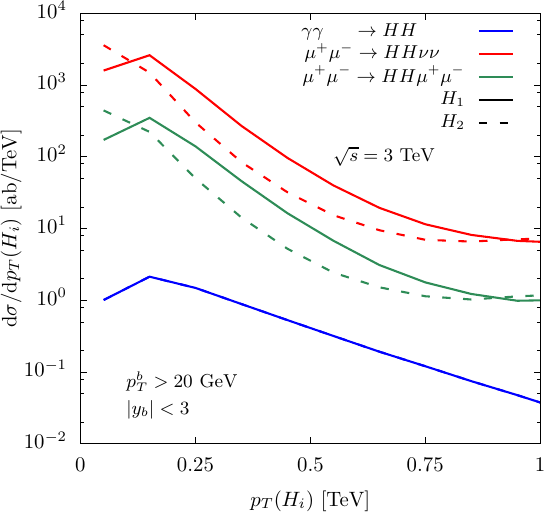}\hskip 0.5 cm\includegraphics[width=0.94\columnwidth]{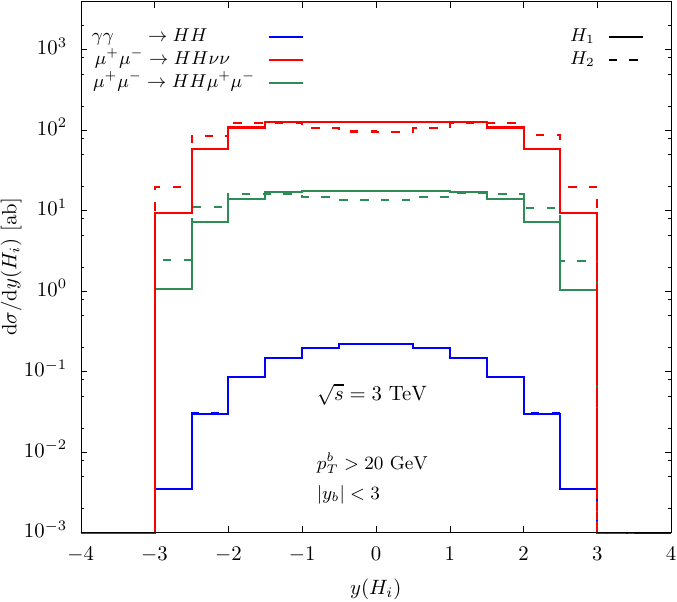}
   \vskip 0.3 cm
%
  \includegraphics[width=0.90\columnwidth]{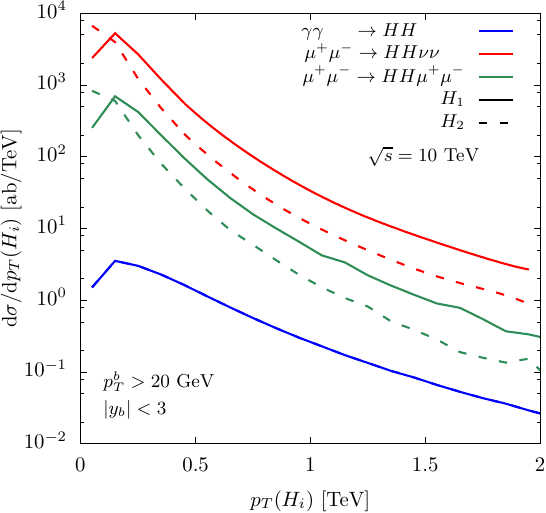}\hskip 0.5 cm\includegraphics[width=0.94\columnwidth]{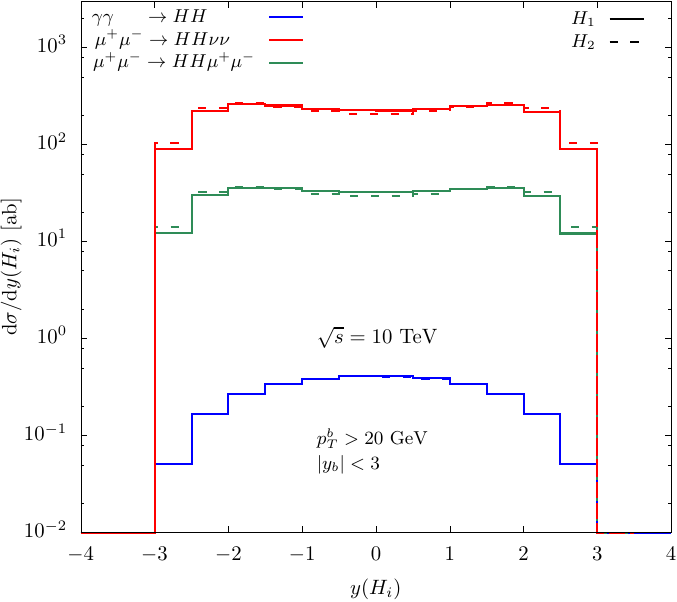}
   \vskip 0.3 cm
%
  \includegraphics[width=0.90\columnwidth]{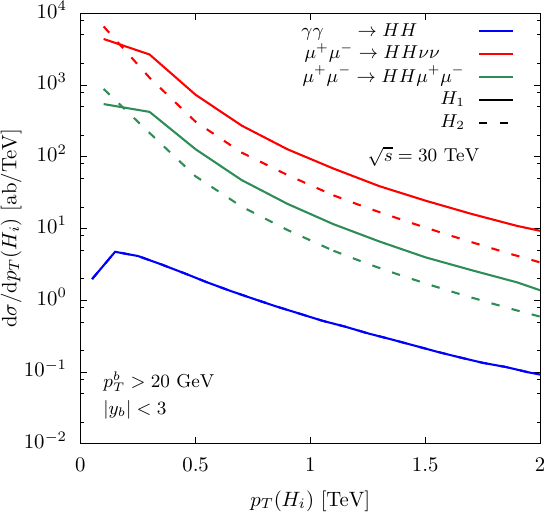}\hskip 0.5 cm\includegraphics[width=0.94\columnwidth]{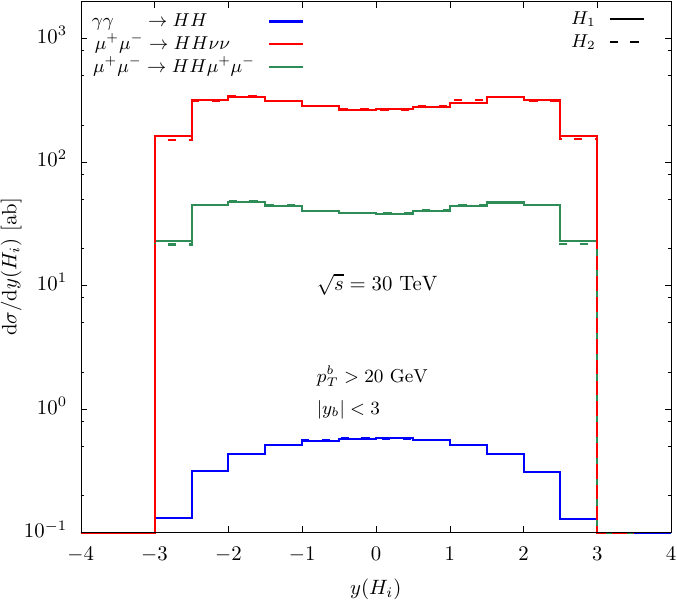}
       \caption{\label{fig:kinHH_acc} 
       Higgs transverse-momentum (left plots) and rapidity (right plots) distributions for the two-Higgs photon-induced production in $\mu^+\mu^- \!\to \mu^+\mu^- HH\,$,   at different $\rs$, after imposing a limited geometric acceptance for the Higgs
     two-body decay products. $H_{1 (2)}$ labels the (second) highest-$p_T$  Higgs.
     The same distributions for the
      $\mu^+\mu^-\to HH\nu \bar{\nu}$ and $\mu^+\mu^-\to HH\mu^+\mu^-$
     tree-level channels   
     are  shown for comparison.}
\end{figure*}

\begin{figure*}
  \includegraphics[width=0.90\columnwidth]{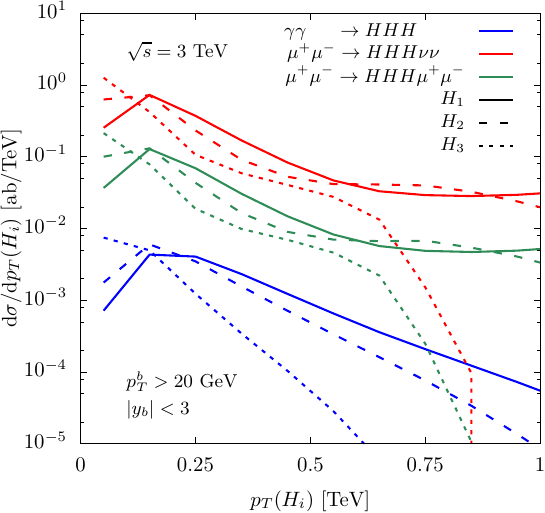}\hskip 0.5 cm\includegraphics[width=0.94\columnwidth]{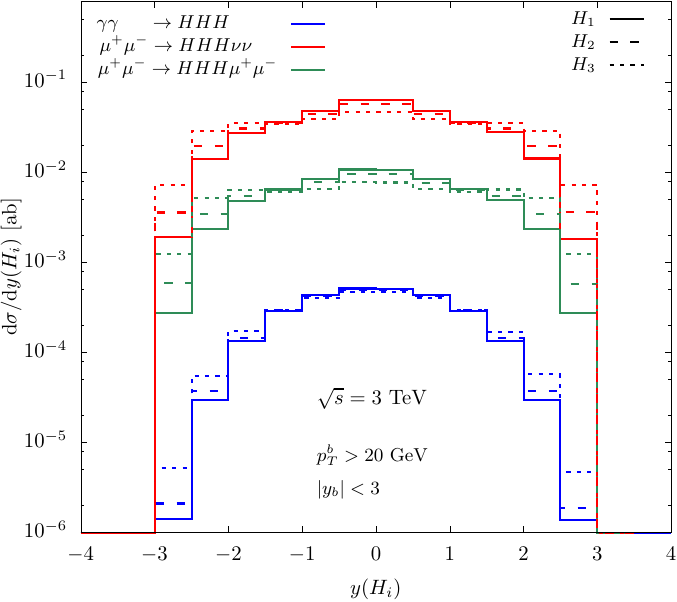}
   \vskip 0.3 cm
%
  \includegraphics[width=0.90\columnwidth]{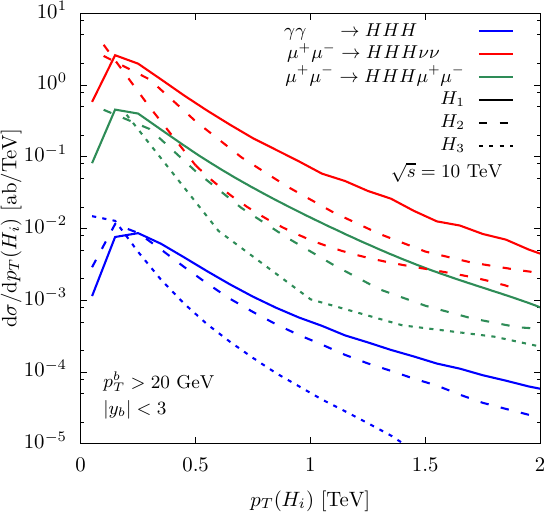}\hskip 0.5 cm\includegraphics[width=0.94\columnwidth]{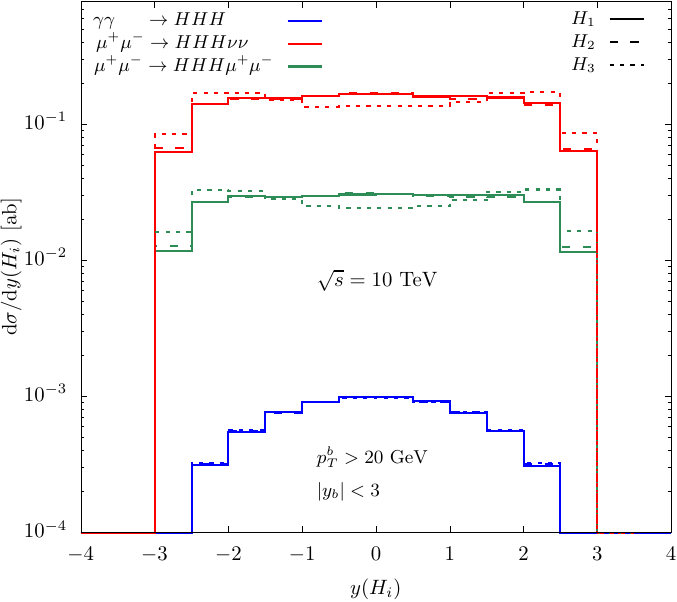}
   \vskip 0.3 cm
%
  \includegraphics[width=0.90\columnwidth]{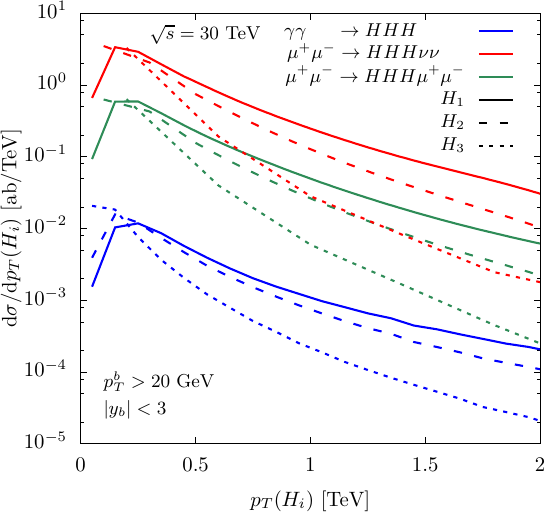}\hskip 0.5 cm\includegraphics[width=0.94\columnwidth]{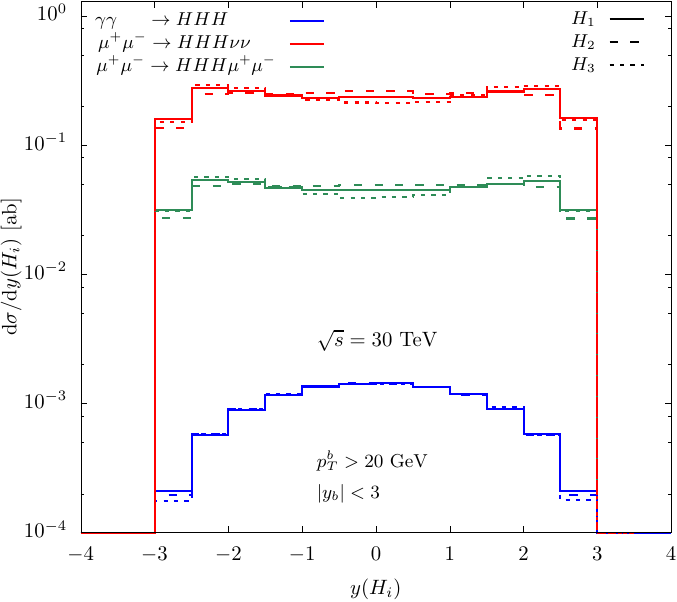}
         \caption{\label{fig:kinHHH_acc}
             Higgs transverse-momentum (left plots) and rapidity (right plots) distributions for the triple-Higgs photon-induced production in $\mu^+\mu^- \!\to \mu^+\mu^- HHH\,$,   at different $\rs$, after imposing a limited geometric acceptance for the Higgs
     two-body decay products. $H_{1 (2/3)}$ labels the (second/third) highest-$p_T$  Higgs.
     The same distributions for the
      $\mu^+\mu^-\to HHH\nu\bar{\nu}$ and $\mu^+\mu^-\to HHH\mu^+\mu^-$
     tree-level channels   
     are  shown for comparison.}
\end{figure*}

In this section, we go through the main kinematical features of the Higgs bosons
in $\gamma\gamma$-induced $\mu^+\mu^- \!\to \mu^+\mu^- (HH,HHH)\,$
production. 
We first present the inclusive Higgs kinematical distributions. 
Then we include the effect of a reduced acceptance in the Higgs kinematics
due to a possible shielding of the detector in 
the forward regions, where the {\it beam-induced background} (BIB) might become  prohibitive~\cite{Cummings:2011zz,Kahn:2011zz,Mokhov:2011zzd,Mokhov:2014hza,DiBenedetto:2018cpy,Bartosik:2019dzq,Bartosik:2020xwr,Lucchesi:2020dku}. To
this end, we will assume a two-body decay of the produced Higgs
(giving the dominant contribution to  the Higgs decay width, through the channels
$H\to b\bar b, c\bar c, gg, \tau\tau$), and put a minimum
transverse-momentum cut and a maximum  rapidity cut on the Higgs decay products (applying no decay branching ratio)
\footnote{ In the following (in particular in Figures~\ref{fig:kinHH_acc},~\ref{fig:kinHHH_acc}, and~\ref{fig:sigma_cut})
we will label the Higgs decay product as $b$ given the importance of the $H\to b\overline{b}$ decay channel,
but the actual flavour of these particles is immaterial since no branching ratio is applied in the calculation. }.
In particular, we simulate the effect of  introducing ``shielding nozzles'' in the detector,
by assuming a detector coverage for all kind of particles
only for transverse momentum and rapidity  $p_T> 20 $ GeV, $|y|<3$.

As for the single Higgs production, 
in the present approximation, the  Higgs  kinematics simply reflects  
the rapidity distribution of the initial $\gamma\gamma$ system as provided by the 
collinear photon spectrum described by  the 
EPA in Eq.~(\ref{eq:EPA}), which gives  vanishing  Higgs
transverse momentum. 

\begin{figure*}
  \includegraphics[width=0.94\columnwidth]{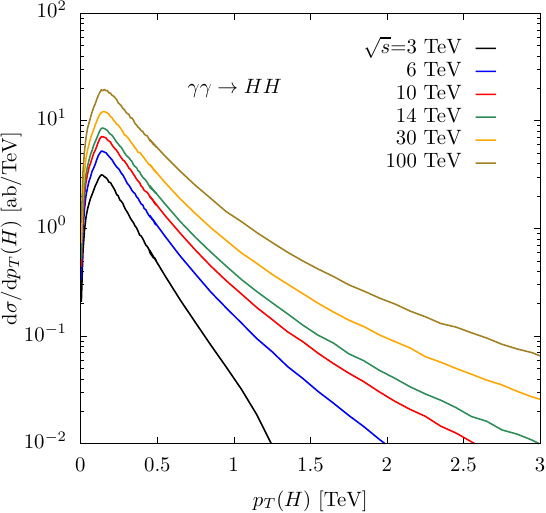}\hskip 0.5 cm\includegraphics[width=0.94\columnwidth]{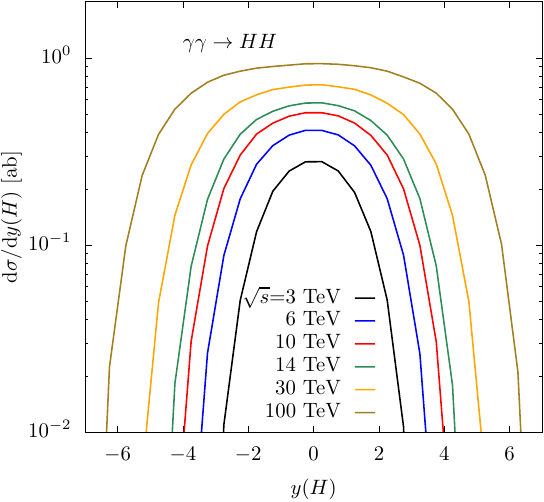}
    \caption{\label{fig:kinHH} 
  Higgs transverse-momentum (left plot) and rapidity (right plot) distributions for the two-Higgs photon-induced production in $\mu^+\mu^- \!\to \mu^+\mu^- HH\,$,   versus $\rs$.
   }
\end{figure*}

\begin{figure*}
  \includegraphics[width=0.94\columnwidth]{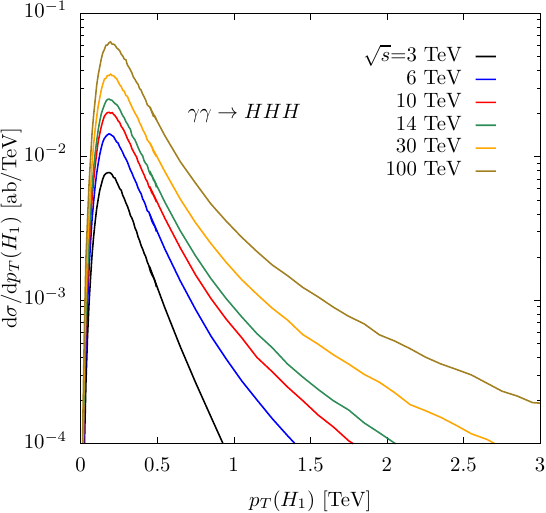}\hskip 0.5 cm\includegraphics[width=0.94\columnwidth]{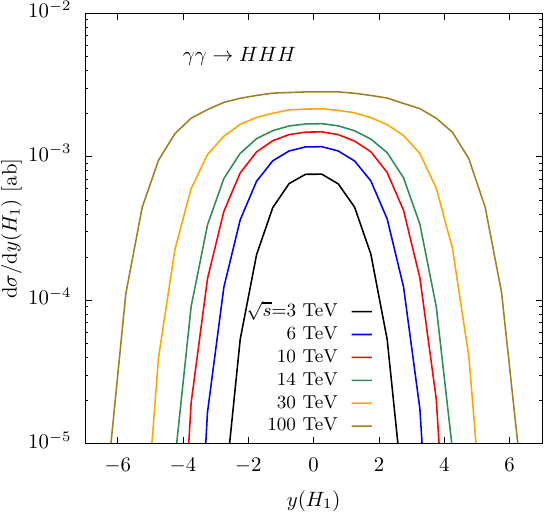}
  \vskip 0.5 cm
%
  \includegraphics[width=0.94\columnwidth]{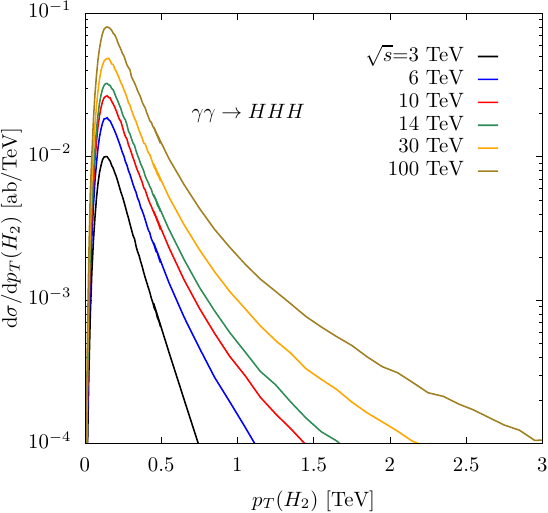}\hskip 0.5 cm\includegraphics[width=0.94\columnwidth]{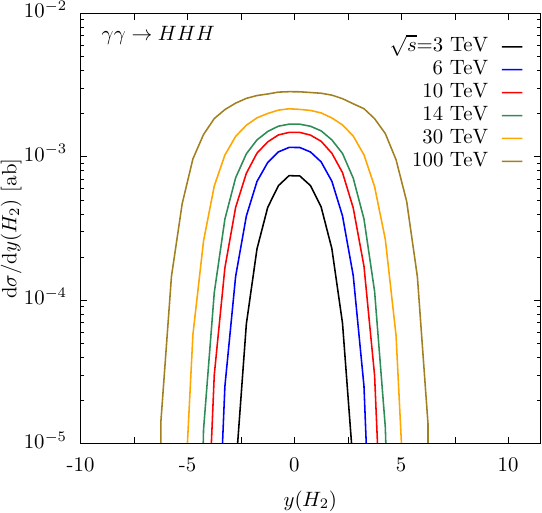}
      \caption{\label{fig:kinHHH} 
  Higgs transverse-momentum (left plots) and rapidity (right plots) distributions for the three-Higgs photon-induced production in $\mu^+\mu^- \!\to \mu^+\mu^- HHH\,$,   versus $\rs$. $H_{1 (2)}$ labels the (second) highest-$p_T$  Higgs.
   }
\end{figure*}

The inclusive Higgs transverse-momentum and rapidity distributions for the two Higgs production in $\mu^+\mu^- \!\to \mu^+\mu^- HH\,$ 
mediated by $\gamma^\ast \gamma^\ast \to HH$  
are presented in
Figure~\ref{fig:kinHH}, for different values of $\rs$.
The same distributions for the three Higgs production in $\mu^+\mu^- \!\to \mu^+\mu^- HHH\,$ 
 mediated by $\gamma^\ast \gamma^\ast \to HHH$  are shown in Figure~\ref{fig:kinHHH}, where
$H_{1}$ labels the highest-$p_T$  Higgs, and $H_{2}$ the second highest-$p_T$  Higgs.

In Figure~\ref{fig:kinHH_acc}, we present instead the
 photon-induced 
$\mu^+\mu^- \!\to \mu^+\mu^- HH\,$
Higgs transverse-momentum and rapidity distributions after applying
the selection cuts  $p_T^b> 20 $ GeV, $|y_b|<3$ 
on the $H\to b \bar b$ decay products (blue lines), at  $\rs=3,10, 30$ TeV.
For comparison, the same distributions for the
 tree-level 
$WW$ channel (red lines)
and $ZZ$ channel (green lines) are also shown. For the process 
$\mu^+\mu^- \!\to \mu^+\mu^- HH\,$, calculated with EPA, both Higgs bosons
have the same transverse momentum and therefore only the blue solid line
is shown. 

Similarly, in Figure~\ref{fig:kinHHH_acc}, we present the 
 $\mu^+\mu^- \!\to \gamma^\ast \gamma^\ast \mu^+\mu^- \!\to \mu^+\mu^- HHH\,$  case. As anticipated, 
no Higgs decay BR is applied in Figure~\ref{fig:kinHH_acc} and \ref{fig:kinHHH_acc}.

\section{Enhancing the $\gamma\gamma\to n\,H$ signal over the
$WW,ZZ\to n\,H$ background}
The previous discussion clearly shows that one-loop $\gamma\gamma\to H,HH,HHH$ production in $\mm$ collisions is quite suppressed with respect
to the $WW,ZZ\to H,HH,HHH$ production. Independently of that, 
 a $\gamma\gamma\to HHH$ signal can be considered statistically out of reach even for the most 
 optimistic luminosities scenarios. On the other hand, the higher statistics
 channels $\gamma\gamma\to H,HH$, being suppressed by orders of magnitudes with respect to $WW,ZZ\to H,HH$ production, will be in general overwhelmed
 by the latter, 
making the detection of loop-induced events quite hard in general.
In this section we propose a strategy that could dramatically enhance a  
$\gamma\gamma\to H,HH,HHH$ signal over the VBF ones.

It is well know that VBF single Higgs production is characterized by
a typical Higgs transverse momentum of the order of the $M_{W,Z}$
[Eq.~(\ref{ptVBF})].
  This feature  
is shared by VBF {\it multiple} Higgs production, and is  connected to the mechanism of vector-boson radiation 
 from the initial beams.
 
\begin{figure*}
  \includegraphics[width=0.94\columnwidth]{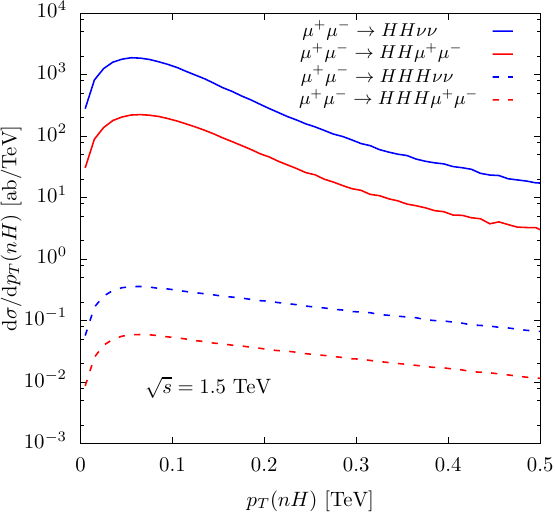}\hskip 0.4 cm\includegraphics[width=0.94\columnwidth]{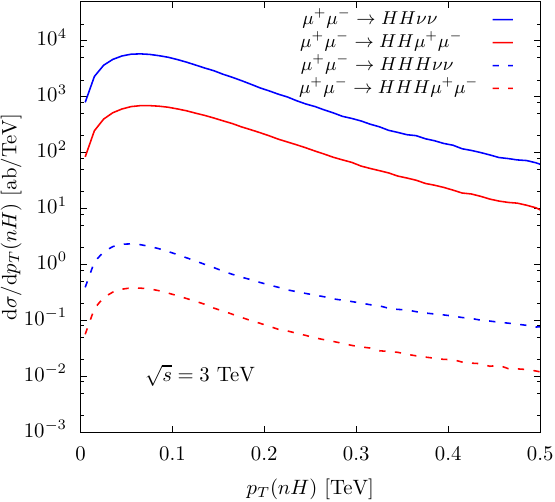}
   \vskip 0.3 cm
%
  \includegraphics[width=0.94\columnwidth]{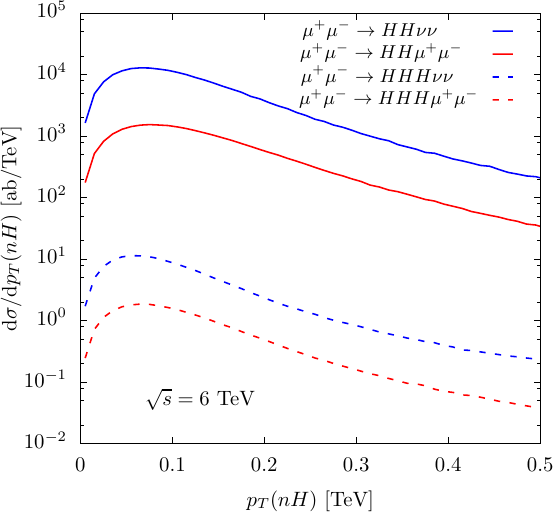}\hskip 0.4 cm\includegraphics[width=0.94\columnwidth]{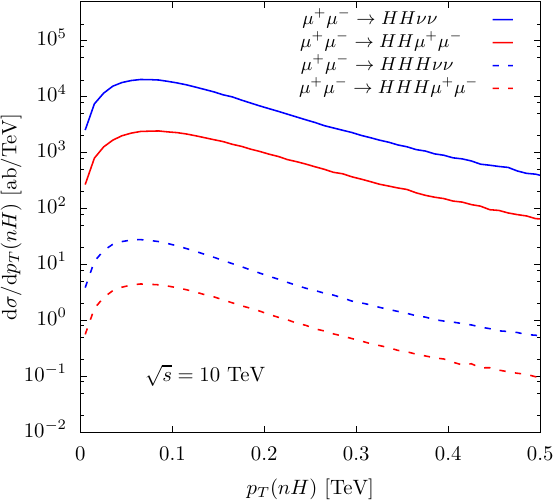}
   \vskip 0.3 cm
%
  \includegraphics[width=0.94\columnwidth]{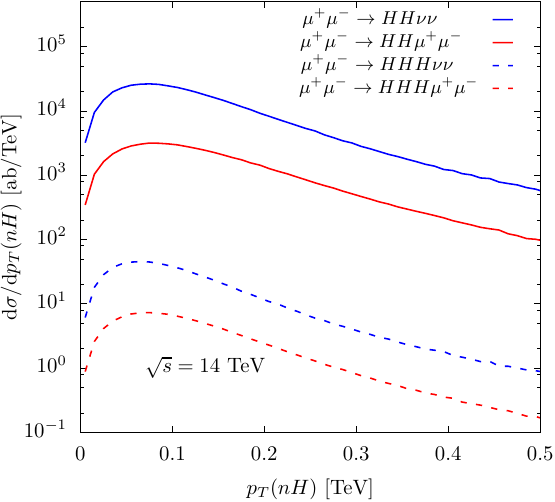}\hskip 0.4 cm\includegraphics[width=0.94\columnwidth]{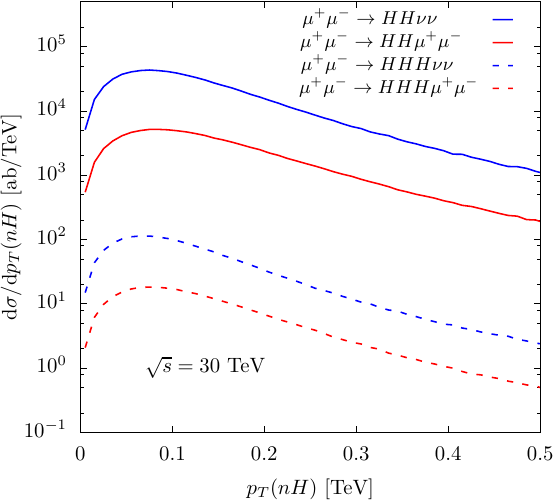}
       \caption{\label{fig:pTSb}
       Transverse-momentum distribution of the $HH$ (solid lines) and $HHH$ (dashed lines) systems in vector-boson-fusion mediated double and triple Higgs production, respectively, in muon collisions, at different $\rs$.
   }
\end{figure*}

 In Figure~\ref{fig:pTSb}, we show 
    the transverse-momentum distribution of the $HH$ (solid lines) and $HHH$ (dashed lines) systems in VBF-mediated $HH$ and $HHH$  production, respectively, in $\mm$ collisions, at various $\rs$.
       The  transverse-momentum distribution for the 
       $HH$ and $HHH$ systems indeed keeps the single Higgs VBF production main feature,   with a maximum at $p_T(HH,HHH)\sim M_{W,Z}$.
       This should be confronted with the $\gamma\gamma$ one-loop process
       where, in the approximation of Eq.~(\ref{eq:EPA}), one has 
       $p_T(H,HH,HHH)\sim 0$.
       
       In order to enhance the $\gamma\gamma$ signal over the VBF one, one can then impose,   depending on the momentum resolution $\Delta p_T(n\,H)$ of the detector,  a cut on large  $p_T(n\,H)$
       events (that is events characterized by $p_T(n\,H)\gsim \Delta p_T(n\,H)$), which leaves unsuppressed  the photon-induced 
       $p_T(n\, H)\sim 0$ events.
       In this way, the actual background from VBF events is reduced 
       to the sub-dominant fraction of events characterized by a small $p_T(n\, H)$, hopefully as small as much less than ${\cal O}(M_W)$.

\begin{figure*}
  \includegraphics[width=0.94\columnwidth]{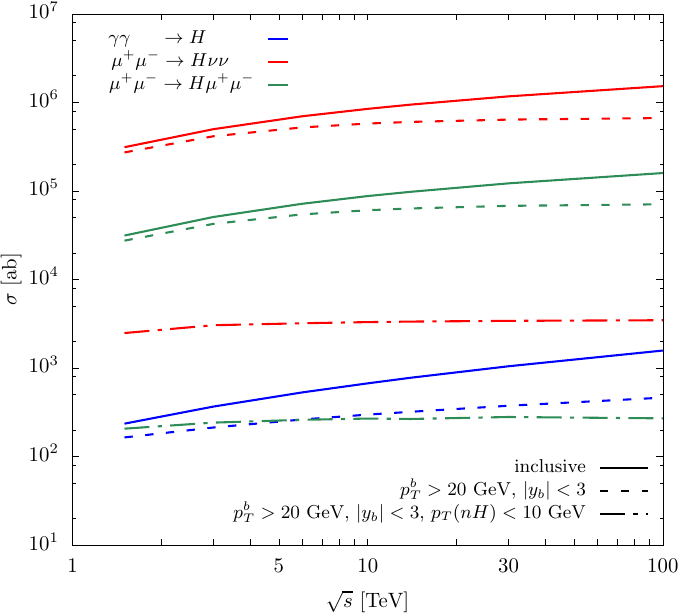}\hskip 0.5 cm \includegraphics[width=0.94\columnwidth]{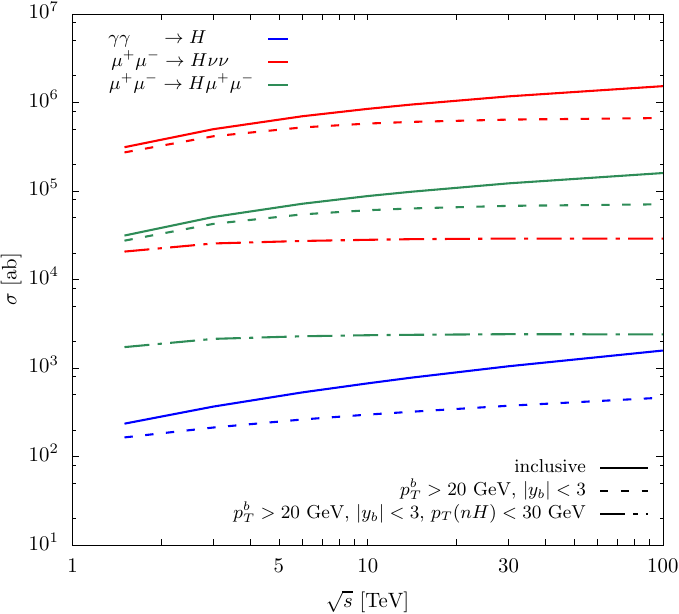}
   \vskip 0.3 cm
%
  \includegraphics[width=0.94\columnwidth]{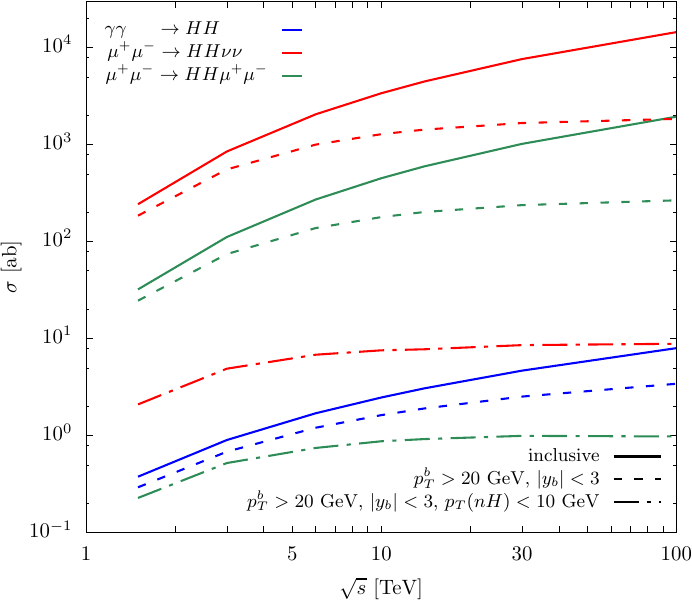}\hskip 0.5 cm \includegraphics[width=0.94\columnwidth]{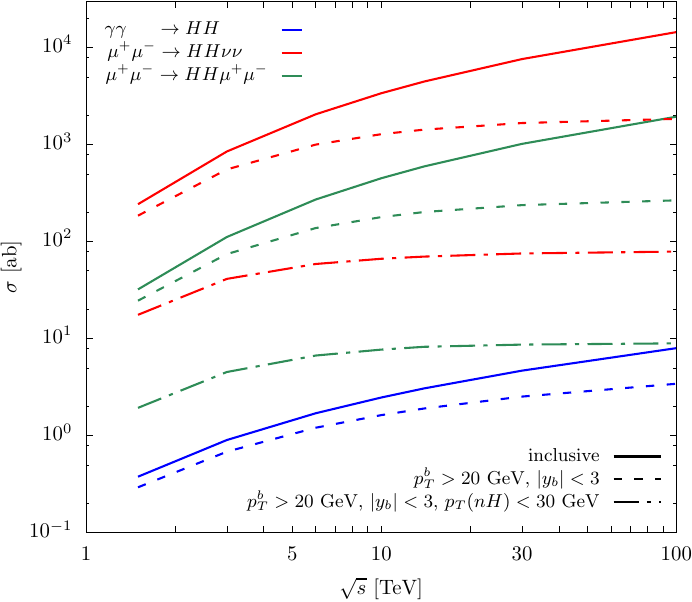}
   \vskip 0.3 cm
%
  \includegraphics[width=0.94\columnwidth]{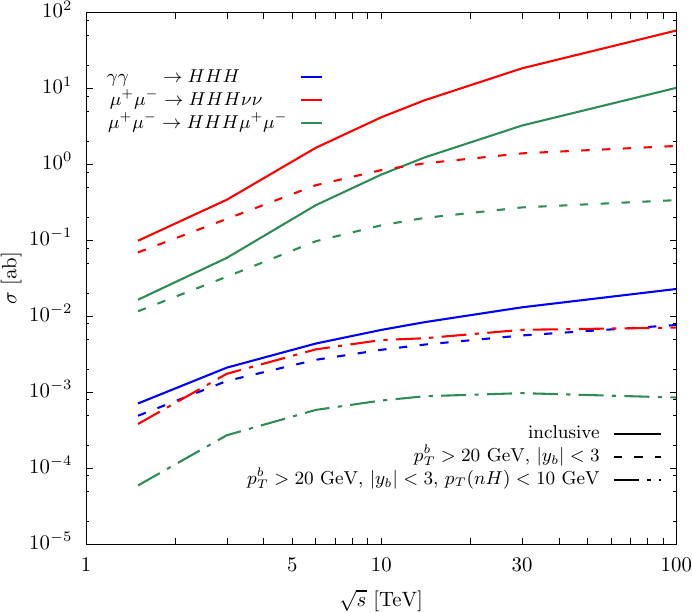}\hskip 0.5 cm \includegraphics[width=0.94\columnwidth]{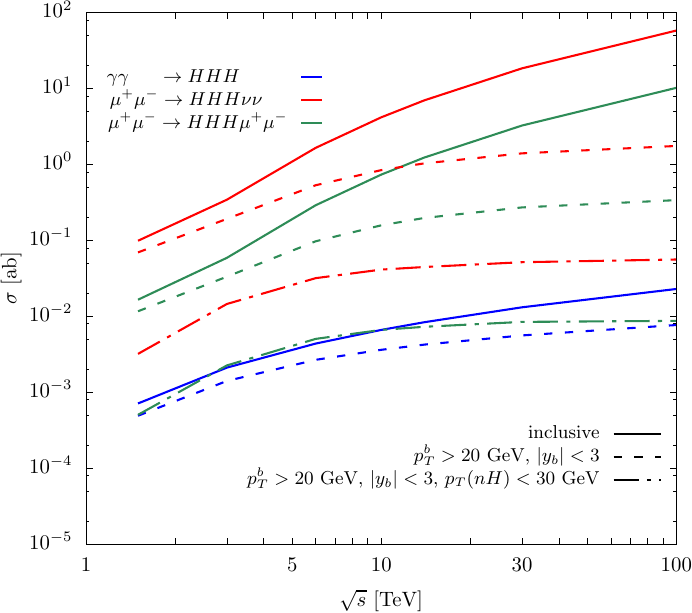}
           \caption{\label{fig:sigma_cut}
           Effects of the cut-flow selecting  low $n\,H$ transverse-momentum events.
       The $\mu^+\mu^-$ inclusive cross section for the $n\,H$ processes 
       (solid lines) are compared with the production rates that pass the Higgs decays basic acceptance (dashed lines),
       and with the rates where a further constrain on the maximum $n\,H$  transverse-momentum (as reconstructed from Higgs decay products) is applied  (dot-dashed lines), which is 10 GeV in the left plot, and 30 GeV in the right plot.
   }
\end{figure*}

  %
Figure~\ref{fig:sigma_cut} shows 
           the effects on $\mu^+\mu^-$ total rates of a cut-flow selecting  low $n\,H$ transverse-momentum events.
     The $n\,H$ transverse-momenta are reconstructed by 
       summing up the  vectorial  momenta of the $H\to b \bar b$ decay products,
       assuming the finite acceptance $p_T^b> 20$~GeV, $|y_b|<3$ (not including any decay BR) .       
       Then, the   inclusive cross sections for the $n\,H$ processes 
       (solid lines) are compared with the production rates that pass the $H\to b \bar b$ basic acceptance cuts (dashed lines),
       and with the rates where a further constrain on the maximum $n\,H$  transverse-momentum (as reconstructed from the Higgs decay products) is applied  (dot-dashed lines). Two cases are analyzed for  
       the  maximum $n\,H$  transverse-momentum,
       which are $\;p_T(n\,H)\lsim$10 GeV (left plot), and 
       30 GeV (right plot).

Figure~\ref{fig:sigma_cut} clearly shows how a cut on large transverse momentum of the Higgs system can upset the hierarchy of the rates for the $\gamma\gamma$ and VBF processes, hence allowing to control
the VBF background to multiple Higgs $\gamma\gamma$ production.
In particular, in the $\;p_T(n\,H)\lsim$10 GeV case (left plots), one can see
that after the acceptance cuts the $\gamma\gamma$ 
single Higgs production gets comparable
  to the $ZZ$ channel at all $\rs$, while both the double and triple Higgs $\gamma\gamma$ productions get the upper-hand over the $ZZ$ channels.
  The $WW$ channel is maintained  moderately dominant over $\gamma\gamma$ for both $H$ and $HH$ productions, and gets comparable to $\gamma\gamma$ in $HHH$
  production.
  The VBF background suppression gets of course  less dramatic
  by relaxing the $\;p_T(n\,H)$ to 30~GeV (Figure~\ref{fig:sigma_cut},
  right plots).

   The presence of a veto on the total transverse momentum of the $nH$ system clearly affects also the differential distributions
of the background studied in Figures~\ref{fig:kinHH_acc} and~\ref{fig:kinHHH_acc}. On the one hand, the main effect of such cut
is to induce a suppression of the $\mu^+\mu^-\to HH(H)\nu\bar{\nu}$ and $\mu^+\mu^-\to HH(H)\mu^+\mu^-$ processes similar to the
one observed at the integrated cross-section level. On the other hand, the veto on $p_T(nH)$ also introduces some shape distortions
for the background processes, in particular for the Higgs-boson $p_T$ distributions where the impact of this cut becomes more severe
for increasing Higgs $p_T$ values.  

\begin{figure*}
  \includegraphics[width=0.94\columnwidth]{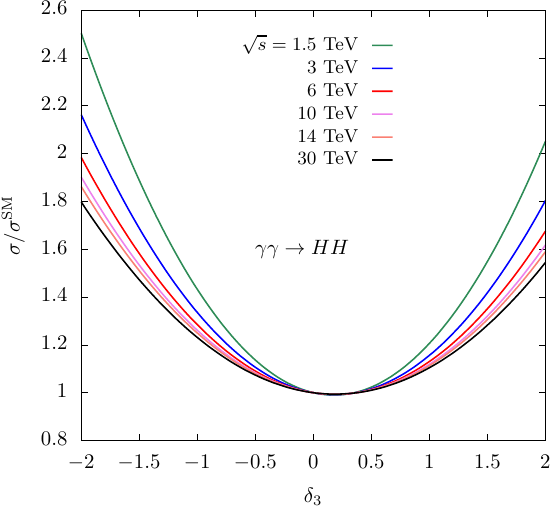}\hskip 0.3 cm
  \includegraphics[width=0.94\columnwidth]{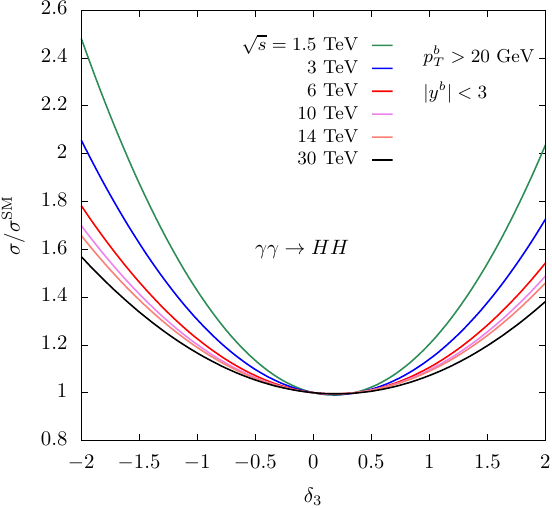}
  \caption{\label{fig:delta_HH}  Ratio of the integrated cross sections with and without anomalous Higgs self-couplings 
    for the process $\gamma\gamma\to HH$ in the inclusive setup (left plot) and after imposing
    acceptance cuts on the Higgs decay products (right plot) as a function of $\delta_3$.
   }
\end{figure*}

\begin{figure*}
  \includegraphics[width=0.94\columnwidth]{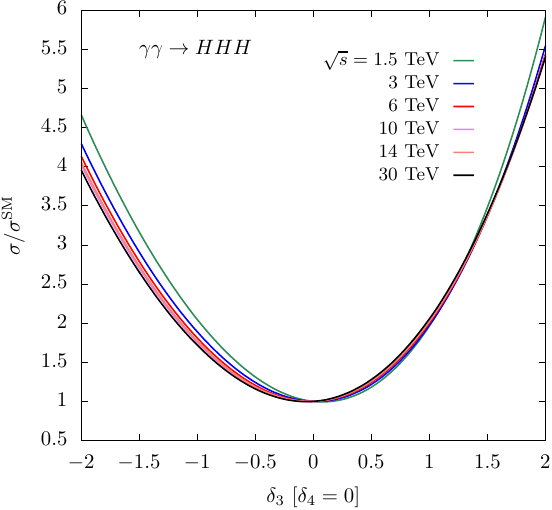}\hskip 0.3 cm
  \includegraphics[width=0.94\columnwidth]{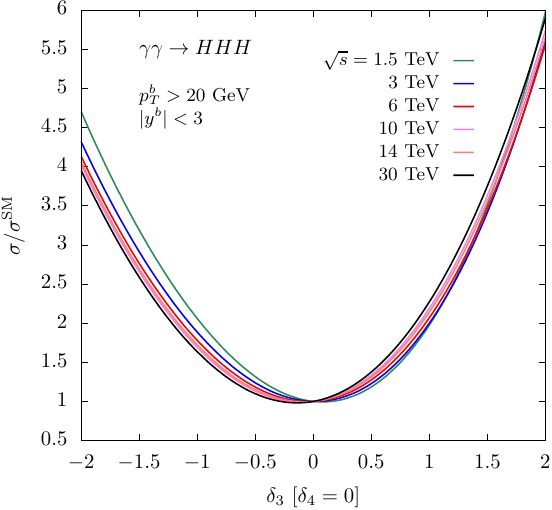}
  \vskip 0.3 cm
  \includegraphics[width=0.94\columnwidth]{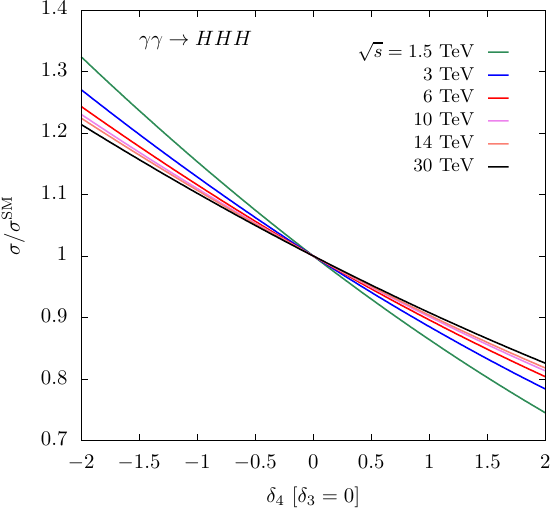}\hskip 0.3 cm
  \includegraphics[width=0.94\columnwidth]{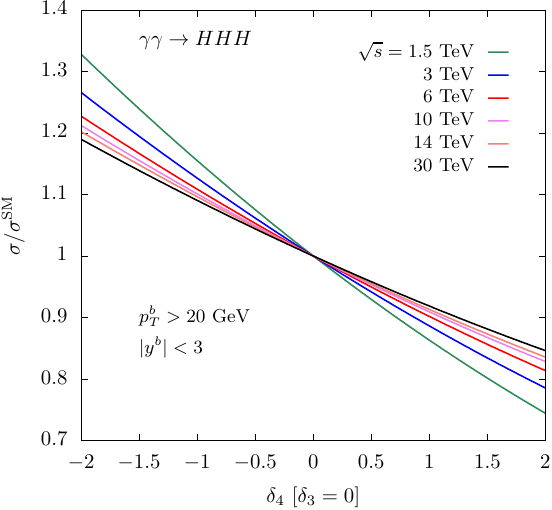}
  \caption{\label{fig:delta_HHH}   Ratio of the integrated cross  sections with and without anomalous Higgs self-couplings 
    for the process $\gamma\gamma\to HHH$ in the inclusive setup (left plots) and after imposing
    acceptance cuts on the Higgs decay products (right plots) as a function of $\delta_3$ (upper plots) and $\delta_4$ (lower plots).
   }
\end{figure*}
\section{Cross section sensitivities to  anomalous  Higgs self-couplings}
Before concluding, we want to briefly  discuss the sensitivity of the 
$\gamma\gamma\to HH,HHH$ cross sections to possible anomalies
in the Higgs boson self-interactions. In particular, we will assume
a deviation in either the trilinear Higgs coupling $\lambda_3 (1+\delta_3) HHH$ or the quartic Higgs coupling $\lambda_4 (1+\delta_4) HHHH$,
where $\lambda_{3,4}$ are the self-coupling SM values, and  $\delta_{3,4}\neq 0$ would 
reveal possible deviations BSM in the Higgs potential.
 We leave a more sophisticated analysis performed in terms of effective-field theories for future work.  
In Figure~\ref{fig:delta_HH}, left plot, we detail the
 ratio of the $\gamma\gamma\to HH$ total cross section with and without anomalous Higgs self-couplings  versus $\delta_3$
at different muon collision $\rs$, while, in the right plot, we show the same quantity
after applying the acceptance cuts on Higgs decays. At the level of inclusive quantities, there is a good sensitivity to anomalous
trilinear coupling of the double Higgs production.
In particular, a $\delta_3\simeq \pm 1$ variation increases the cross sections
by about 15--20\%.  After imposing the acceptance cuts, the sensitivity is in general slightly reduced

In Figure~\ref{fig:delta_HHH} we show the $\delta_3$ dependence
for $\delta_4=0$ (upper plots), and $\delta_4$ dependence
for $\delta_3=0$ (lower plots) for  the ratio of the  cross sections for triple Higgs production with and without anomalous couplings  
via photon scattering at different muon $\rs$. The results corresponding to the setups with (right plots) and without (left plots) acceptance cuts are shown.   
Note that for the reference integrated luminosities in Eq.~(\ref{lumi}), 
the tiny $\gamma\gamma\to HHH$ production cross section
will not allow a precision cross-section measurement. However,
as in the case of double Higgs production, there is a sizeable dependence on anomalous Higgs self-couplings, in particular  for coupling variations of the order of $\delta_{3(4)}\sim \pm 1$, at $\delta_{4(3)}\simeq 0$. Furthermore, $\gamma\gamma\to HHH$ cross section 
 has a minimum around its SM value for $\delta_4\sim0$ and variable 
 $\delta_3$, while the $\delta_4$ dependence is monotonic in the 
 $\delta_4$ range considered in Figure~\ref{fig:delta_HHH} (lower plots).
  We checked that the latter monotonic $\delta_4$ dependence is the outcome of a nontrivial interplay between the top-loop and the $W$-loop amplitudes  providing the final sensitivity to anomalous~$\delta_4$.


\section{Conclusions}
Single, double and triple Higgs-boson production mediated by photon fusion via loops of heavy particles (top quarks and $W$ bosons) have been studied at multi-TeV
lepton colliders. For integrated collision luminosities scaling as 
${\cal L}\sim  10$ 
${\rm ab^{-1}}(\rs/10\,{\rm TeV})^2$, at $\rs \sim 14 (3)$ TeV
one has inclusive cross sections corresponding to 
16.000 (400) single Higgs events and 60 (1) double Higgs events, while
triple Higgs production is quite suppressed, and  needs $\rs \gsim 30$~TeV for getting more than 1 $HHH$ event.
We have studied kinematical distributions at different $\rs$, and 
analyzed the effects of a finite acceptance on the Higgs decay
products in order to control BIB effects via dead cones along the beams. 
The comparison with the dominant VBF mechanism for producing  single, double and triple Higgs final states has been carried out,
and possible strategies for pinpointing $\gamma\gamma$-induced events
on the basis of the total transverse momentum of the Higgs system
have been suggested.
Finally, the inclusive cross sections sensitivity to triple and quartic Higgs self-couplings have been discussed and found sizeable.

Multi-TeV muon colliders offer a unique possibility to probe such processes.
Apart from single Higgs production, which has been discussed at
future $e^+e^-$ colliders~\cite{dEnterria:2017jmj}, we are not aware of any similar study
at multi-TeV future colliders~\cite{Abada:2019lih} allowing to compare the corresponding potential for multi-Higgs production induced by
quasi-real photon collisions.

Our study proceeded under the strict assumption of collinear
initial photons described by means of the EPA,
which is valid  in first approximation.
A more refined analysis should relax this assumption,
and consider  in $\gamma\gamma$ one-loop multi-Higgs
production also interference and coherence effects
with $Z$-boson mediated processes described in terms
of electroweak PDFs~\cite{Chen:2016wkt}.
This would imply  including  at the amplitude level both VBF $ZZ$
fusion channels (possibly including higher-order EW corrections) and one-loop $\gamma Z$-induced channels.
However, the effects of the latter two processes should affect mainly
non-vanishing $p_T(nH)$ final states, and the description in terms
of collinear EW PDFs might need to be improved to better describe
the transverse momenta of the initial-state vector bosons.
We then leave an extensive discussion of this issue to further dedicated studies.

\vspace{10pt}

\begin{acknowledgements}  
We thank Fabio Maltoni for useful discussions. 
\end{acknowledgements}

\bibliographystyle{spphys}       
\interlinepenalty=10000
\bibliography{nH}

\end{document}